\documentclass[a4paper,11pt]{article}
\usepackage[hmargin={2truecm,2truecm},vmargin={1.5truecm,2.5truecm}]{geometry}
\usepackage{amsmath}
\usepackage{amssymb}
\usepackage{graphics} % recomendado en samples
\usepackage{epsfig}
\usepackage{multicol}
\usepackage{multirow}
\usepackage{longtable}
\usepackage{hyperref}
\usepackage{color,soul}
\usepackage{cases}
\usepackage{graphicx}
\usepackage{caption}
\usepackage{subcaption}
\usepackage{authblk}
\usepackage{mathtools}
\usepackage{setspace}
\usepackage{tikz}
\usepackage{arydshln}
\usepackage{rotating}
\usepackage{hhline}
\usepackage{booktabs}
\usepackage{adjustbox}%%%%
\usepackage{tabularx}%%%
\usepackage{colortbl}
\usepackage[normalem]{ulem} % to strikethrough
\usepackage[round]{natbib}
\PassOptionsToPackage{
        natbib=true,
        style=authoryear-comp,
        hyperref=true,
        backend=biber,
        maxbibnames=99,
        firstinits=true,
        uniquename=init,
        maxcitenames=2,
        parentracker=true,
        url=false,
        doi=false,
        isbn=false,
        eprint=false,
        backref=true,
            }   {biblatex}
\newtheorem{thm}{Theorem}

\begin{document}

\title{Managing ESG Ratings Disagreement in Sustainable Portfolio Selection}
\author{Francesco Cesarone$^1$, Manuel Luis Martino$^1$, Federica Ricca$^{2}$, Andrea Scozzari$^3$\\
\normalsize $^1$ Universit\`{a} Roma Tre: \normalsize \{francesco.cesarone, manuelluis.martino\}@uniroma3.it\\
\normalsize $^2$ Sapienza Universit\`{a} di Roma, Italy: \normalsize federica.ricca@uniroma1.it\\
\normalsize $^3$ Universit\`{a} degli Studi Niccol\`{o} Cusano Roma, Italy: \normalsize  andrea.scozzari@unicusano.it\\
}
\date{\today}
\maketitle

\maketitle

\begin{abstract}
Sustainable Investing identifies the approach of investors whose aim is twofold: on the one hand, they want to achieve the best compromise between portfolio risk and return, but they also want to take into account the sustainability of their investment, assessed through some Environmental, Social, and Governance (ESG) criteria. The inclusion of sustainable goals in the portfolio selection process may have an actual impact on financial portfolio performance. ESG indices provided by the rating agencies are generally considered good proxies for the performance in sustainability of an investment, as well as, appropriate measures for Socially Responsible Investments (SRI) in the market. In this framework of analysis, the lack of alignment between ratings provided by different agencies is a crucial issue that inevitably undermines the robustness and reliability of these evaluation measures. In fact, the ESG rating disagreement may produce conflicting information, implying a difficulty for the investor in the portfolio ESG evaluation. This may cause underestimation or overestimation of the market opportunities for a sustainable investment.\\
In this paper, we deal with a multi-criteria portfolio selection problem taking into account risk, return, and ESG criteria. For the ESG evaluation of the securities in the market, we consider more than one agency and propose a new approach to overcome the problem related to the disagreement between the ESG ratings by different agencies. We propose a nonlinear optimization model for our three-criteria portfolio selection problem. We show that it can be reformulated as an equivalent convex quadratic program by exploiting a technique known in the literature as the $k-$sum optimization strategy.
An extensive empirical analysis of the performance of this model is provided on real-world financial data sets.
\end{abstract}


%%%%%%%%%%%%%%%%%%%%%%%%%%%%%%%%%%%%%%%%%%%%%%%%%%%%%%%%%%%%%%%%%%%%%%%%%%%%%%%%%%%%%%%%%%%%%%%%%%%%%%%%%%%%%%%%%%%%%%%%%%%%%%%%
\section{Introduction}\label{sec1-Intro}

Over the past two decades, many studies have focused on the Environmental, Social, and Governance (ESG) criteria when choosing appropriate investment strategies.
The growing interest in sustainable investing as an academic research topic has been accompanied by considerable interest from the financial industry and companies, too.
According to the latest Principles for Responsible Investment (PRI) report \cite{PRI2023}, a total of 5.372 investors representing more than 121.3 trillion of dollars in terms of AUM (Asset Under Management), have committed to integrate ESG information into their investment decisions.
The idea is that investors want to achieve a good trade-off between risk and return on investment, but simultaneously they want to support ethical corporate behaviors.
The interest in sustainable investments is also motivated by the recent EU tendency to make sustainability evaluations an integral part of its financial policy to support the European Green Deal.
In the EU's context, sustainable finance is understood as finance to support economic growth while reducing pressures on the environment and taking into account social and governance aspects.
In fact, the European Union strongly encourages the transition to a low-carbon, more resource-efficient, and sustainable economy to build a financial system for a sustainable growth.
In this context, the aim of Sustainable Investing (SI) is to combine ESG evaluation aspects with the standard investment criteria adopted in portfolio selection.
Two main strategies can be applied when choosing investment opportunities according to ESG evaluations.
On the one hand, one can follow a two-phase approach, trying to reduce the dimension of the financial market to a smaller subset of assets satisfying suitable ESG requirements and then considering only these assets in the portfolio construction \citep[see, e.g.,][and the references therein]{liagkouras2020incorporating}.
On the other hand, one could try to include ESG aspects in the financial analysis for the selection process.
For a comprehensive review on socially responsible investing and sustainability indicators, the reader is referred to the very recent review by \cite{koenigsmarck2023shifting}.

\noindent
Over the years, the literature has brought attention to the relationship between ESG ratings and the financial performance of an investment \citep[see, e.g.,][]{derwall2005eco,friede2015esg,brooks2018effects,bermejo2021esg}.
Most of the papers agree that sustainability has a positive impact on financial performances \citep[][]{amon2021passive,cesarone2022does}, and there is evidence that companies that include sustainability criteria in their decisions tend to outperform those that do not \citep[][]{nofsinger2014socially,khan2016corporate,kolbel2017media}.
Despite the amount of research on sustainable investments, we observe that the results depend closely on the ESG evaluation criteria used by the major agencies.
In this regard, ESG rating providers have become influential institutions (see, e.g., \cite{berg2022aggregate}), but they are often characterized by some lack of common definitions of ESG features and attributes when setting ESG standards.
Indeed, ESG ratings provided by different agencies typically differ substantially and often produce disagreement on the evaluation of the same company \citep[see, e.g.,][]{berg2022aggregate,billio2021inside,gibson2021esg,chatterji2016ratings}.
As a consequence, disagreement over ESG ratings could generate conflicting information for investors, potentially underestimating or overestimating market opportunities for sustainable investments.

\noindent
This paper deals with a portfolio selection problem including ESG evaluation criteria.
Focusing on the above-illustrated agency disagreement problem, the novelty in our approach is a strategy that takes into account, at the same time, all the ESG scores that the different agencies assign to the same firm.
Our method is able to overcome the problem of the lack of coherence in the different ESG scores given by the many agencies in the market but still considering all agencies' evaluations simultaneously.
In fact, all papers presented in the recent literature which incorporate ESG criteria in portfolio construction, are able to consider only one agency evaluation, implying the issue of which agency has to be chosen among the many.
For the selected agency, different traditional and widely known models for portfolio selection have been applied to construct ESG portfolios, such as the Markowitz mean-variance portfolio optimization approach \citep{cesarone2022does,utz2014tri,utz2015tri,steuer2023non}.
Likewise, more recent portfolio optimization methods have been proposed for the construction of ESG portfolios based on the CVaR minimization as in \cite{morelli2023responsible}.
Another line of research is based on the Bayesian optimization technique applied for maximizing the performance of a portfolio of stocks under the presence of ESG criteria incorporated into the objective function. A recent and comprehensive description of the state of the art in portfolio optimization with ESG criteria can be found in \cite{garrido2023bayesian}.

\noindent
Our methodology applies a more sophisticated screening phase in which all the ESG evaluations assigned to the securities by the agencies are considered simultaneously in a concise ESG measure, which becomes part of the objective function of our portfolio optimization model.
We exploit some results developed for the $k-$sum optimization problem \citep[see, e.g.,][]{puerto2017revisiting,punnen1992k,punnen1996k}.
The $k-$sum optimization strategy is an extension of the standard min-sum optimization where only the $k$ largest cost coefficients of the objective function are included in the sum.
In our financial problem we have to select a portfolio of stocks among $n$ risky assets by considering a vector of $m$ ESG scores assigned by $m$ agencies to the $n$ assets.
More precisely, we model a three-criteria portfolio selection problem, to find a portfolio whose variance is minimized while the expected return and the (concise) ESG score are maximized.
The resulting model is a quadratic $k-$sum optimization model. We show that, by exploiting the results in \cite{bertsimas2003robust} and \cite{puerto2017revisiting}, we can reformulate our model as a convex quadratic program where the objective function incorporates the variance of portfolio returns, the expected return, and the ESG portfolio evaluation measure.

\noindent To summarize, in the following, we list the three main contributions of this paper.
From an application perspective of portfolio selection, we develop a methodology that provides two new operational tools:
(1) we introduce a new general method to handle the multiplicity of ESG scores assigned to securities in the market by different and independent rating agencies;
(2) we develop a strategy to simultaneously tackle the problem of screening the assets in the market based on their ESG performance while selecting an optimal portfolio, thus solving our multiple-criteria optimization problem in a single step.
From a theoretical viewpoint, we show how the class of $k-$sum optimization problems, already introduced in the literature for linear programs, can be extended to the quadratic case (3).

In practice, we use this last result to solve our three-criteria convex quadratic programming problem and analyze the performance of the Pareto-optimal portfolios found.
We then show how the same results can be exploited to perform the empirical construction of the efficient frontier of our problem in three dimensions.
To obtain such approximated frontiers, we apply an $\epsilon$-constraint approach consisting of a convex quadratic programming model where the portfolio variance is minimized by imposing parametric bounds on the required levels of portfolio expected return and ESG score.
Our empirical analysis, performed on two real-world data sets from major stock markets, provides insights into the effect of considering ESG ratings in portfolio selection and the positive impact that this strategy can have on the out-of-sample portfolio performance in terms of risk and return. These results are compared with other standard portfolio selection methods, as well as, with the portfolio optimization strategy based on a single agency evaluation provided in \citep{cesarone2022does}, and the results show that our new method outperforms the others.

\noindent
The rest of the paper is organized as follows.
Section \ref{sec:Disagreement} introduces stylized facts about ESG disagreement.
We describe the different metrics used by important rating agencies to assess the sustainability of companies and we introduce a scaling technique to map each (original) ESG rating into a (new) common scale.
In Section \ref{sec:TheoFramework} we present the theoretical framework and we introduce a new general method to manage the ESG rating disagreement in portfolio selection.
We show how to reformulate our nonlinear portfolio optimization problem as a convex quadratic programming by exploiting the $k-$sum optimization strategy \citep[][]{puerto2017revisiting,ponce2018mathematical}.
Then, we describe how to obtain the efficient surface of the corresponding three-criteria portfolio selection model.
In Section \ref{sec:EmpAnalysis}, we provide an out-of-sample performance analysis of the proposed approach basing on some real-world financial data sets.
Finally, Section \ref{sec:Conclusions} draws some conclusions and emphasizes the importance of our approach in the current financial framework.

%%%%%%%%%%%%%%%%%%%%%%%%%%%%%%%%%%%%%%%%%%%%%%%%%%%%%%%%%%%%%%%%%%%%%%%%%%%%%%%%%%%%%%%%%%%%%%%%%%%%%%%%%%%%%%%%%%%%%%%%%%%%%%%%
\section{On the disagreement of the ESG scores}\label{sec:Disagreement}

In this section, we first describe the main characteristics of the ESG scores retrieved from the different rating agencies considered.
Then, we present the scaling technique used to guarantee that the ESG scores of different agencies provide coherent measures of the same aspect.
Finally, we provide some stylized facts that highlight the disagreement of the ESG scores in the real markets.

\noindent In Table \ref{tab:DataProviders_Summary} we list four among the most important and commonly used data providers \citep[see, e.g.,][]{gibson2021esg}, and, for each of them, we specify the meaning of the extremal values of the score ranges (the \emph{Brownest} or the \emph{Greenest}).

\begin{table}[htbp!]
	\centering
	\scalebox{0.9}{\begin{tabular}{cccccccccccccc}
			\toprule
			Rating Provider & & Abbreviation && Score Range & & Metric\\
			\midrule
			\text{Refinitiv} & & RFT & & 0 (Brownest)\,-\,100 (Greenest) & & ESG Score\\
            \text{Bloomberg} & & BMG & & 0 (Brownest)\,-\,100 (Greenest) & & ESG Disclosure Score\\
            \text{Morningstar Sustainalytics} & & MNG & & 0 (Greenest)\,-\,100 (Brownest) & & ESG Risk Score\\
            \text{S\&P Global} & & S\&P & & 0 (Brownest)\,-\,100 (Greenest) & & ESG Global Rank\\
			\bottomrule
	\end{tabular}}
        \caption{ESG Data Providers}
        \label{tab:DataProviders_Summary}
\end{table}

\noindent
Refinitiv's ESG Scores are designed to comprehensively and objectively measure a company's ESG performance, engagement and effectiveness, based on verifiable company-reported data in the public domain.
ESG scores are data-driven and take into account the most relevant industry metrics, with minimal company size and transparency biases.
Refinitiv's ESG scoring methodology follows several key calculation principles.
Refinitiv collects ESG data across the globe by various metrics, looking at annual reports, news sources, Corporate Social Responsibility (CSR) reports, stock exchange filings, company websites, feeding a proprietary ESG database.
The information collected is mapped into 10 main categories including emissions, environmental product innovation, human rights, shareholders and so on, which together define the three pillar scores: Environmental (E), Social (S) and Governance (G).
Each category has a specific weight, which is calculated based on the Refinitiv magnitude matrix, and which varies per industry for the environmental and social categories, while remains the same across all industries for the governance categories.
The resulting overall ESG score is the weighted sum of the 10 category scores, and ranges between 0 and 100, indicating the lowest and highest ESG performance, respectively (see \cite{ESGRefinitiv}).

\noindent
Bloomberg's ESG Disclosure Score measures the amount of ESG data that each company publicly discloses across the three pillars of environmental, social and governance (see \cite{tamimi2017transparency}).
Bloomberg evaluates the completeness of reporting using 120 ESG qualitative and quantitative indicators such as carbon emissions, diversity, shareholders rights and so on, which are tailored to different industry sectors.
Bloomberg searches a wide variety of public documents and sources through which companies disclose ESG information, that include CSR or sustainability reports, company websites, policy-related reports, direct communication (e.g. surveys), press releases, third-party research, and news items.
The overall ESG Disclosure Score combines the three pillars using a proprietary method, and is published annually on a scale from 0 for companies that do not publish any ESG data included in the pillars, to 100 for those that meet the completeness requirements.

\noindent
Morningstar Sustainalytics's ESG Risk Score is based on the concept of risk decomposition, to derive the level of unmanaged risk for a company \citep[see, e.g.,][]{ESGSustainalytics,christensen2022corporate}.
Companies typically engage in programs, practices, policies, and actions to manage ESG risks. This score measures to which extent a company is able to manage this risk.
The score ranges from 0 and 100, with 0 indicating that ESG risks have been fully managed and 100 indicating the highest level of unmanaged risk.

\noindent
S$\&$P Global Rank's ESG Score measures a company's performance in terms of ESG risks management, where the risk is considered material if it presents a significant impact on society or the environment and a significant impact on a company's value drivers, competitive position, and long-term shareholder value creation (see \cite{ESGS&PGLOBAL}).
It combines business information, media and stakeholder analysis, modeling techniques, and company engagement.
According to S$\&$P Global's methodology, ESG Scores are measured on a scale of 0-100, where 100 indicates the maximum score, and reflect the company's performance on ESG topics.
Clearly, since ESG issues tend to be industry-specific, the company's overall sustainability performance is compared with its peers within its industry.
%
%These question-level scores aggregate up to the criteria-level, which reflect the most concrete ESG topics (depending on the sub-industry).
%
%The criteria-level scores further combine to form standalone E, S and G scores, which ultimately roll up into a single, overall ESG score for every company.
%

\noindent
Although all of these metrics range between 0 and 100, three of them assign higher scores to better ESG performance (i.e., Refinitiv, Bloomberg, and S$\&$P Global), whereas Morningstar Sustainalytics associates higher scores with worse ESG risk management (i.e., the lower the score, the greener the company is).
%
%To address this issue, we do the following.
To address this problem, the first step is to standardize the data.

\medskip

Assume that there are $n$ risky assets and $m$ different rating agencies and that each agency evaluates the sustainability of each asset according to its own method and producing its ESG measure, which is a real number ranging between a minimum and a maximum value.
Even if the sustainability principles on which these measures are based are the same for all agencies, depending on each agency methodology, the same asset typically has different evaluations by two different agencies.

To guarantee that ESG evaluations by different providers are coherent and given in the same range, we perform a feature scaling of the ESG different measures \citep[see, e.g.,][]{jain2011min,pandey2017comparative,sinsomboonthong2022performance}.

\noindent
Let $e_{ij}$ be the ESG score that agency $i$ assigns to asset $j$, with $i=1,\dots,m$ and $j=1,\dots,n$.
Since for all scores the range of possible values is a close and bounded interval, we first normalize the different ESG scores as follows:
\begin{equation*}
\overline{e}_{ij} = \displaystyle\frac{e_{ij}-e_{j}^{min}}{e_{j}^{max}-e_{j}^{min}}
\end{equation*}
where $e_{i}^{min}=\displaystyle\min_{1 \leq j \leq n} e_{ij}$ and $e_{i}^{max}=\displaystyle\max_{1 \leq j \leq n} e_{ij}$.
Clearly, the scaled ESG score $\overline{e}_{ij}\in[0,1]$.
%
%\begin{equation*}
%\overline{e}_{ij} = \displaystyle\frac{e_{ij}-e_{ij}^{min}}{e_{ij}^{max}-e_{ij}^{min}}
%\end{equation*}
%
%where the (new) scale defined by the minimum ESG score $e_{ij}^{min}$ and the maximum ESG score $e_{ij}^{max}$ of a data set is such that $\overline{e}_{ij}\in[0,1]$.
%
%We observe that the min-max normalization seems to represent an effective choice, since ESG scores are measured on a bounded scale.
%

\noindent
We then define the \emph{Non-ESG score} $s_{ij}\in[0,1]$ as the complement of the normalized ESG score, namely $s_{ij}=1-\overline{e}_{ij}$.
Hence, the lower $s_{ij}$ the greener the asset $j$ for agency $i$.
We denote by $s^{i}$ the column vector of the Non-ESG scores assigned to the $n$ assets by agency $i$, with $i=1,\dots,m$.
To simplify the presentation, when this does not cause any confusion, we refer to these numbers as ESG \emph{scores}, even if we know that they are Non-ESG scores.

\medskip

\begin{figure}[htbp!]
        \centering
   \includegraphics[width=1\linewidth]{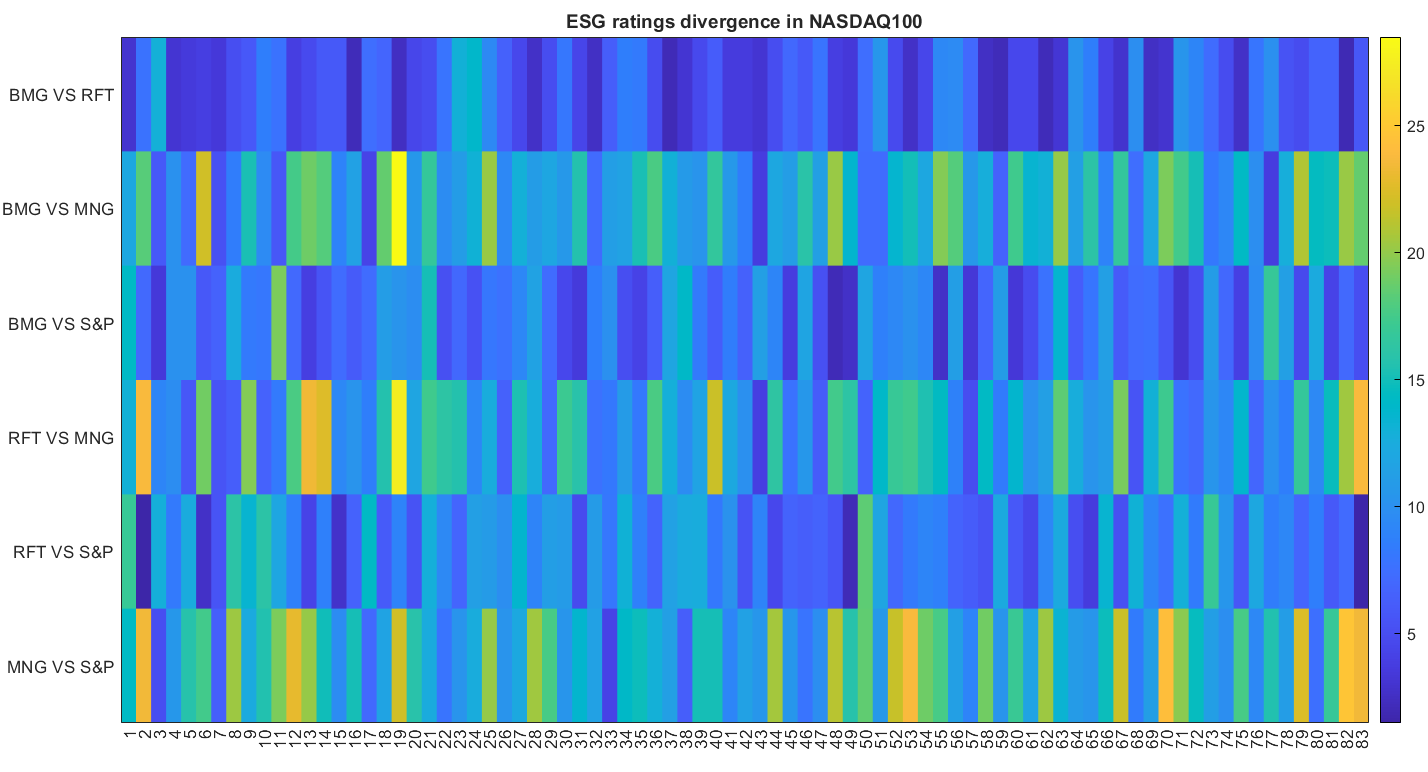}
    \caption{ESG ratings divergence among data providers in the NASDAQ100 financial market}
 \label{fig:ESG_Disagreement_NASDAQ100}
\end{figure}

\begin{figure}[htbp!]
        \centering
	\includegraphics[width=1\linewidth]{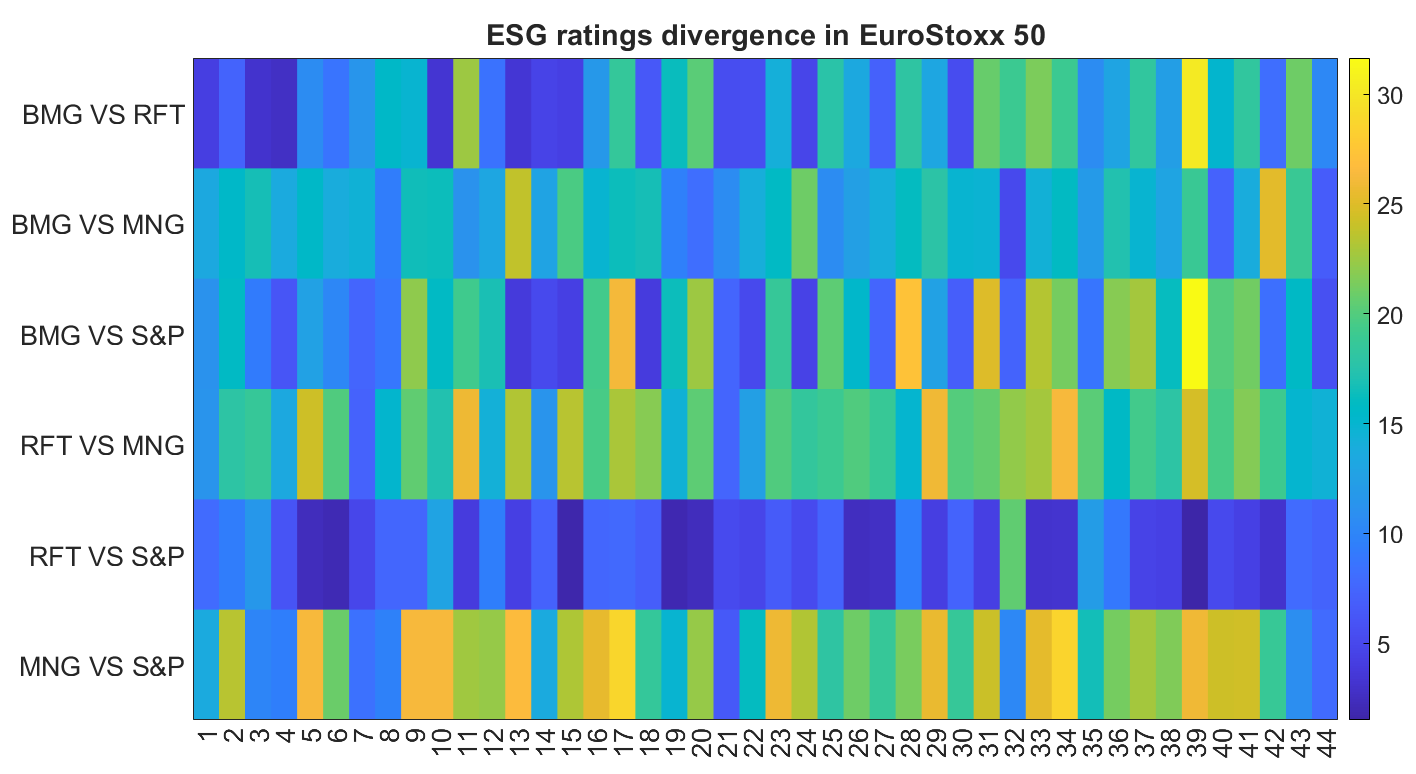}
	\caption{ESG ratings divergence among data providers in the EuroStoxx50 financial market}
 \label{fig:ESG_Disagreement_EUROSTOXX50}
\end{figure}

Once all scores are mapped to a common scale, the ratings of the different data agencies can be compared.
To quantify the ESG disagreement, for each asset we compute the pairwise distances between ESG scores assigned by different agencies.
Figures \ref{fig:ESG_Disagreement_NASDAQ100} and \ref{fig:ESG_Disagreement_EUROSTOXX50} provide evidence of the pairwise comparison for the NASDAQ100 and EuroStoxx50 financial markets, respectively. On the horizontal axis we report each asset of the considered market, while on the vertical axis we consider the six possible comparisons.
In Figure \ref{fig:ESG_Disagreement_NASDAQ100} we observe misalignments of the ESG scores between any pair of agencies for the NASDAQ100 market, especially when Morningstar Sustainalytics (MNG) is involved. A similar situation is reported in Figure \ref{fig:ESG_Disagreement_EUROSTOXX50} for EuroStoxx50.

%
% In Figures \ref{fig:ESG_Disagreement_NASDAQ100} and \ref{fig:ESG_Disagreement_EuroStoxx50} we illustrate the ESG ratings assigned by the data providers listed in Table \ref{tab:DataProviders_Summary} to the constituents of the EuroStoxx 50 and the NASDAQ100 stock markets, respectively.
%
% The asset constituents of each market are reported on the x-axis.
% %
% Clearly, the ratings assigned by the data providers to each asset show up visually as vertical lines.
%

\noindent
Furthermore, to understand whether there is some systematic differences between the NASDAQ100 and the EuroStoxx50 markets, in Table \ref{tab:Divergences} we also provide the average (euclidean, chebychev, cosine, correlation) distances of the scores over all assets and over all pairs of data providers.

% For the purpose of our empirical analysis, we computed the average divergence in the ESG ratings assigned by all the rating providers (see Table \ref{tab:RatingProviders}) to all the constituents of the stock markets listed in Table \ref{tab:DailyDatasets}, using several distance metrics.
%
\begin{table}[htbp!]
	%\small
	\centering
	%\begin{center}
	\scalebox{0.9}{\begin{tabular}{cccccccccccccc}
			\toprule
			Market Index & & Euclidean & &  Chebychev & & Cosine & & Correlation\\
			\midrule
			\text{NASDAQ100} & & 10.22 & & 0.40 & & 0.05 & & 0.35\\
            \text{EuroStoxx 50} & & 16.00 & & 0.56 & & 0.07 & & 0.75\\
			\bottomrule
	\end{tabular}}
	%\end{center}
        \caption{Average ESG ratings divergence among rating providers}
	\label{tab:Divergences}
\end{table}
\noindent The different values of the distance metrics suggest that the rating providers seem to agree more in their ESG ratings for the US market than those for the European one.

\noindent
These results suggest that our $k-$sum optimization approach will be probably more effective when the markets are characterized by more intense disagreement among rating agencies.

%\begin{itemize}
%
%\item \textbf{Descrizione data provider (chi da brown, chi da green, 0-100, ecc):}
%
%\item \textbf{Normalizzazione dati con motivazione:}
%
%\item \textbf{Disallineamento (tabelle, statistiche descrittive, grafici):}
%
%\end{itemize}

%%%%%%%%%%%%%%%%%%%%%%%%%%%%%%%%%%%%%%%%%%%%%%%%%%%%%%%%%%%%%%%%%%%%%%%%%%%%%%%%%%%%%%%%%%%%%%%%%%%%%%%%%%%%%%%%%%%%%%%%%%%%%%%%
\section{Theoretical Framework}\label{sec:TheoFramework}

We  propose a portfolio selection strategy that follows a risk-return approach but also considers the sustainability of the investment measured through a global ESG index, which summarizes in a single measure the evaluations of all the agencies.
For this aspect, we follow a worst-case analysis approach, according to which we try to limit the inclusion in the portfolio of assets with bad ESG scores.

\noindent
In Section \ref{subsec:k-sum} we introduce the general $k-$sum operator \citep[][]{puerto2017revisiting,ponce2018mathematical}.
In Section \ref{subsec:Models}, we include in the Mean-Variance framework a new objective that takes into account the ESG ratings provided by multiple agencies, and we show how to reformulate the resulting multi-objective nonlinear portfolio optimization problem as a convex quadratic programming, by exploiting the $k-$sum optimization strategy.
Then, in Section \ref{subsec:EffSurface}, we describe the methodology for the empirical construction of the efficient surface of the set of investment opportunities in the given market.

%%%%%%%%%%%%%%%%%%%%%%%%%%%%%%%%%%%%%%%%%%%%%%%%%%%%%%%%%%%%%%%%%%%%%%%%%%%%%%%%%%%%%%%%%%%%%%%%%%%%%%%%%%%%%%%%%%%%%%%%%%%%%%%%
\subsection{Portfolio selection and k-sum optimization}\label{subsec:k-sum}

We consider a setting where $n$ assets are available. The aim of an investor is to select a portfolio composed with a subset of securities that achieves some specified goal. We denote by $x_{j}$, $j=1,\dots,n$, the fraction of capital invested in asset $j$, and assume that short-selling is not allowed.
Therefore, a portfolio $P(x)$ corresponds to a non negative vector $x=(x_{1},x_{2},\dots,x_{n})$ such that $\displaystyle\sum_{j=1}^{n}x_{j}=1$ and $x_{j}\geq 0, j=1, \ldots, n$. The feasible domain of our portfolio selection problem is described by the following set of linear constraints:

\begin{equation*}
	\Omega=\Big\{x\in \mathbb{R}^{n}: \sum_{j=1}^{n}x_{j}=1, x_j \geq 0, j=1,\dots,n\Big\}\,.
\end{equation*}

\noindent For each agency $i$, consider the vector $s^{i}$ of the Non-ESG scores assigned to the $n$ assets by agency $i$ and compute the scalar product between vectors $s^{i}$ and $x$, thus obtaining the expression of the Non-ESG score by agency $i$ for a portfolio $P(x)$:

\begin{equation*}
	s^{i}(x)=(s^{i})^{T}x
\end{equation*}

\noindent Then, for any $P(x)$ we can sort in a non increasing order its $m$ Non-ESG scores assigned by the $m$ agencies:

\begin{equation*}
	s^{\overline \pi_1(x)}(x) \geq s^{\overline \pi_2(x)}(x) \geq \dots  \geq s^{\overline \pi_m(x)}(x)\,,
\end{equation*}

\noindent where

\begin{equation}\label{OrderedScores}
\overline \pi(x)=({\overline \pi_1(x), \overline \pi_2(x), \overline \pi_i(x), \ldots, \overline \pi_m(x)})
\end{equation}

\noindent is the agencies' permutation corresponding to the above score ordering where a lower score corresponds to a greener asset.

%is taken from \cite{filippi2019bridging,puerto2017revisiting,punnen1992k}.

\noindent
We define a global Non-ESG score of a portfolio $x=(x_{1},x_{2},\dots,x_{n})$ as the sum of the $k$ largest (worst) scores in the above ranking, $1 \le k\le m$:

\begin{equation}\label{eq:qWorstOperator}
s_{k}^{\overline{\pi}(x)}(x) = \displaystyle\sum\limits_{i=1}^{k}s^{\overline{\pi}_{i}(x)}(x).%=\max\limits_{\pi_{1}(x),\dots,\pi_{m}(x)}\sum_{i=1}^{q}s^{\pi_{i}(x)}(x)
\end{equation}

\noindent We call this function the $k-$Worst Non-ESG score of $x$. Then, for a given $k$, our investment sustainability objective is formulated as finding a portfolio $x\in \Omega$ that minimizes $s^{\overline \pi(x)}_k(x)$.
If $k=m$, the problem is to find a portfolio $x$ that minimizes the sum of all scores (in fact, in this case, the permutation does not apply).
On the other hand, if $k=1$, the problem is to find a portfolio that minimizes the worst portfolio score among those given by the $m$ agencies.
In all other cases, when $1 < k < m$, we have an intermediate worst-case approach in which we want to consider more than one ESG evaluation, but not all, and, in fact, we consider only the scores of the $k$ agencies that provided the worst evaluations.

A $k-$sum optimization approach is often used to mitigate the effects of data uncertainty in robust optimization (see, e.g. \cite{bertsimas2003robust, BertsimasSim2004}). In fact, it is well known that a minimax model (i.e., $k=1$) provides robust solutions against uncertain data, although it may result too conservative. On the other hand, the minimization of $k$ largest ``looses'', $1 < k < m$, provides more robust solutions than considering just the largest loss ($k=1$), in particular, when the data are correlated as in portfolio selection problems. Actually, in \cite{BertsimasSim2004} the authors show that, in the classical Markowitz framework, the parameter $k$ of a $k-$sum approach for risk minimization represents a sort of protection level of the actual portfolio return.
In any real-world application, parameter $k$ must be suitably chosen according to the available ESG rating agencies and the investor's subjective view of their evaluations.

\vskip 8 pt
\noindent Note that the ordering (\ref{OrderedScores}) establishes the $k$ worst scores to be considered in (\ref{eq:qWorstOperator}) is a function of $x$, implying that, before optimization, it is not known. Therefore, in the model, the objective function is formulated w.r.t. a generic permutation $\pi(x)$, and the resulting portfolio optimization problem is the following:

\begin{equation}\label{Model1}
	\min\limits_{x\in\Omega}\Big\{x^{T}\Sigma x - \mu^{T}x + \max\limits_{\pi_{1}(x),\dots,\pi_{m}(x)}\sum_{i=1}^{k}s^{\pi_{i}(x)}(x)\Big\}\,,
\end{equation}
\noindent where $\Sigma$ and $\mu$ represent the variance/covariance matrix of returns and the vector of expected returns of the given assets, respectively. Note that the last term of the objective function is a maximum over all possible permutations $\pi(x)$.
Solving the above problem is not easy, since finding the optimum depends on a permutation of the scores which is a function of the decision vector $x$.
\noindent
Following the approach in \cite{ponce2018mathematical} and \cite{puerto2017revisiting}, the inner maximization problem in \eqref{Model1} can be re-written, for any given $x$, by introducing $m$ new binary variables $z_{i}$, $i=1,\dots,m$:

\begin{equation}\label{eq:InnerMax}\hspace{-2cm}
	\max\limits_{\pi_{1}(x),\dots,\pi_{m}(x)}\sum_{i=1}^{k}s^{\pi_{i}(x)}(x)=\max\limits_{z}\Big\{\sum_{i=1}^{m}s^{i}(x)z_{i}:\sum_{i=1}^{m}z_{i}=k\,\,\,\,\, z_{i}\in\{0,1\},\, i=1,\dots,m\Big\}.
\end{equation}
Note that when $x$ is fixed the scores $s^{i}(x), i=1,\dots,m$, are constant.
In Problem \eqref{eq:InnerMax}, the feasible vectors $z$ are indicating vectors of the subsets of agencies that have cardinality exactly equal to $k$.
For any given $x$, Problem \eqref{eq:InnerMax} finds the sum of the $k$ largest Non-ESG scores.

\noindent
Since Problem \eqref{eq:InnerMax} is feasible and bounded, and the constraint coefficient matrix is totally unimodular, we can relax the integrality constraints on the $z$ variables into $0\le z_{i}\le 1$, thus obtaining the following linear program equivalent to \eqref{eq:InnerMax}:

\begin{equation}\label{eq:InnerMax-LP}
\max\limits_{\pi_{1}(x),\dots,\pi_{m}(x)}\sum_{i=1}^{k}s^{\pi_{i}(x)}(x)=\max\limits_{z}\Big\{\sum_{i=1}^{m}s^{i}(x)z_{i}:\sum_{i=1}^{m}z_{i}=k\,\quad 0\le z_{i}\le 1,\,\quad i=1,\dots,m\Big\}\,.
\end{equation}
Since $x$ is fixed, \eqref{eq:InnerMax-LP} is a linear program with $m+1$ constraints and $m$ non-negative variables.
For the same fixed $x$, we compute its dual problem in the $m+1$ dual variables, denoted by $u$ and $v_i$ with $i=1,2, \ldots, m$:
\begin{equation}\label{eq:InnerMax-LP-Dual}
\max\limits_{\pi_{1}(x),\dots,\pi_{m}(x)}\sum_{i=1}^{k}s^{\pi_{i}(x)}(x)=\min\limits_{u,v}\Big\{ku + \sum_{i=1}^{m} v_{i}: v_{i}+u\ge s^{i}(x),\, i=1,\dots,m,\, v_{i}\ge 0,\, i=1,\dots,m,\, u\in\mathbb{R}\Big\}\,.
\end{equation}
Since the Non-ESG score vectors $s^{i}$ are all non-negative, in the maximization model \eqref{eq:InnerMax-LP} we can replace $\displaystyle\sum_{i=1}^{m}z_{i}=k$ by $\displaystyle\sum_{i=1}^{m}z_{i}\le k$.
In fact, for any optimal solution of \eqref{eq:InnerMax-LP}, this constraint is always active. Therefore, the above dual model \eqref{eq:InnerMax-LP-Dual} can be equivalently re-written by setting $u\ge 0$.

We can now formulate our original portfolio selection problem \eqref{Model1} as the following mathematical program:

\begin{equation}\label{SingleObj}
		\begin{array}{llll}
            \min\limits_{(x,v,u)} & x^{T} \Sigma x -\mu^{T}x + \Big[ku+\displaystyle\sum_{i=1}^{m}v_{i}\Big] &\\[5pt]
            \mbox{s.t.} & & & \\[5pt]
            & v_{i}+u\ge (s^{i})^{T}x & i=1,\dots,m \\[5pt]
            & v_{i}\ge 0 & i=1,\dots,m \\[5pt]
            & u \geq 0 & & \\[5pt]
\\
            & \displaystyle \sum_{k=1}^{n}x_{j}=1 & \\[5pt]
            & x_{j}\ge 0 & j=1,\dots,n & \\[5pt]
		\end{array}
\end{equation}

\noindent The domain of the above optimization model is, in fact, given by the domain of model (\ref{eq:InnerMax-LP-Dual}) and $\Omega$.
Model \eqref{SingleObj} is a Convex Quadratic Program which can be efficiently solved by applying standard optimization solvers (Cplex, Gurobi, etc.). In Section \ref{sec:EmpAnalysis} we apply this single-objective model in a rolling time-windows framework to generate a sequence of in-sample optimal portfolios for which we then evaluate the out-of-sample performance.

%%%%%%%%%%%%%%%%%%%%%%%%%%%%%%%%%%%%%%%%%%%%%%%%%%%%%%%%%%%%%%%%%%%%%%%%%%%%%%%%%%%%%%%%%%%%%%%%%%%%%%%%%%%%%%%%%%%%%%%%%%%%%%%%

\subsection{Three-objective portfolio optimization model}\label{subsec:Models}

In this paper, we are also interested in evaluating the relationship between the three considered criteria by analyzing the trade-off between the risk, return and ESG performance of a portfolio $x\in \Omega$. Hence, we further investigate on this issue by computing the efficient portfolios w.r.t. the three criteria considered. In this regard, the corresponding three-objective portfolio optimization problem is:

\medskip
\begin{equation}\label{eq:Triobjective}
		\begin{array}{lll}
            \displaystyle\min\limits_{x} & \sigma_{P}^{2}(x)=x^{T}\,\Sigma\,x & \\[8pt]
            \displaystyle\max\limits_{x} & \mu_{P}(x)=\mu^{T} x & \\[8pt]
            \displaystyle\min\limits_{x} & \max\limits_{\pi_{1}(x),\dots,\pi_{m}(x)}\displaystyle\sum\limits_{i=1}^{k}s^{\pi_{i}(x)}(x)& \\[8pt]
            \mbox{s.t.} & & \\[4pt]
            & \displaystyle x\in\Omega &
		\end{array}
\end{equation}
\noindent
Let us consider the function $\max\limits_{\pi_{1}(x),\dots,\pi_{m}(x)}\sum_{i=1}^{k}s^{\pi_{i}(x)}(x)$, that computes the sum of the $k$ largest (nonnegative) elements of a vector ordered in non-increasing order. This function belongs to the class of \emph{Ordered Median Functions} \cite{NickelPuerto2009}. In fact, let us now consider an ordering $\pi_{\leq}(x)$ where $s_{\pi_1(x)}(x) \leq s_{\pi_2(x)}(x) \leq \dots  \leq s_{\pi_m(x)}(x)\,$ are the scores sorted in non-decreasing order. The above function can be written:

\begin{equation}\label{OMF_1}
\max\limits_{\pi_{1}(x),\dots,\pi_{m}(x)}\sum_{i=1}^{k}s^{\pi_{i}(x)}(x)= \displaystyle\sum\limits_{i=m-k+1}^{m}s_{{\pi}_{i}(x)}(x).
\end{equation}

\noindent that is equivalent to

\begin{equation}\label{OMF_2}
\displaystyle\sum\limits_{i=m-k+1}^{m}s^{{\pi}_{i}(x)}(x)=\sum\limits_{i=1}^{m}\alpha_i s_{{\pi}_{i}(x)}(x).
\end{equation}

\noindent where the vector $\alpha$ has components $\alpha_1=\alpha_2=\ldots=\alpha_{m-q}=0$ and $\alpha_{m-q+1}=\alpha_{m-q+2}=\ldots=\alpha_m=1$. Function (\ref{OMF_2}) is the $q$-centrum of $x$. Ordered Median Functions are nonlinear functions. Whereas the nonlinearity is induced by the sorting. Given an Ordered Median Function, if (and only if) the vector $\alpha$ is such that $0\leq\alpha_1,\ldots, \leq\alpha_m$, then the function is convex \cite{NickelPuerto2009}.

\vskip 8 pt
\noindent Since $f_1(x)=\sigma_{P}^{2}(x),f_2(x)=\mu_{P}(x),f_3(x)=\max\limits_{\pi_{1}(x),\dots,\pi_{m}(x)}\sum_{i=1}^{k}s^{\pi_{i}(x)}(x)$ are convex functions, the three-objective optimization problem \eqref{eq:Triobjective} can be solved by applying a weighted sum scalarization of the type $\displaystyle\min\limits_{x\in\Omega}\sum\limits_{h=1}^{3}\lambda_{h}f_{h}(x)$, with $\lambda_{h}\geq 0$ \citep[see Proposition 3.10][]{ehrgott2005multicriteria}, and, to find all the efficient solutions of Problem \eqref{eq:Triobjective}, we have to solve the following model:

\medskip
\begin{equation}\label{eq:Triobjective_lambda}
	\min\limits_{x\in\Omega}\left\{\lambda_{1} \sigma_{P}^{2}(x)  -\lambda_{2}  \mu_{P}(x)  +\lambda_{3}  \left(\max\limits_{\pi_{1}(x),\dots,\pi_{m}(x)}\sum_{i=1}^{k}s^{\pi_{i}(x)}(x)\right) \right\}\,.
\end{equation}
\medskip

\noindent
Exploiting the dual model (\ref{eq:InnerMax-LP-Dual}) introduced in Section \ref{subsec:k-sum}, for any choice of three scalars $\lambda_{h}\ge 0,\, h=1,2,3$, we can formulate a single objective scalar model to find an efficient portfolio of the three-criteria model \eqref{eq:Triobjective}.

\begin{equation}\label{eq:Triobjective_lambda_scalarized}
		\begin{array}{llll}
            \min\limits_{(x,v,u)} & \lambda_1 \sigma_{P}^{2}(x) - \lambda_2 \mu_{P}(x) + \lambda_3 \left(qu+\displaystyle\sum_{i=1}^{m}v_{i}\right) &\\[5pt]
            \mbox{s.t.} & & & \\[5pt]
            & v_{i}+u\ge (s^{i})^{T}x & i=1,\dots,m \\[5pt]
            & v_{i}\ge 0 & i=1,\dots,m \\[5pt]
            & u \geq 0 & & \\[5pt]
            & \displaystyle x\in\Omega &
		\end{array}
\end{equation}

\noindent It is well-known that, for a general multi-criteria optimization problem, there exists a relationship between the optimal solutions of the scalar-sum model \eqref{eq:Triobjective_lambda_scalarized} and those obtained by optimizing one of the three objectives alone, while controlling the value of the other two by bounding constraints. This is known as the $\epsilon$-\emph{constraint approach} approach \citep[see Theorem 4.6 in][]{ehrgott2005multicriteria}. This result is valid under mild assumptions, which hold also in our case. We report this results in the following theorem, which refers to a generic multi-objective minimization problem of the form: $\min\{f_1(x),\ldots,f_p(x)\; | x\in \mathcal{X}\}$ \citep[see also][]{pozo2023biobjective}. Here $R^p_{\geq}$ denotes the set of nonnegative vectors in $R^p$.

\begin{thm} \label{Theorem_1}
\citep[][]{chankong1983optimization,ehrgott2005multicriteria}.
\begin{itemize}
\item[1.] Suppose that $\hat{x}$ is an optimal solution of $\min\limits_{x\in \mathcal{X}} \sum\limits_{h=1}^p \lambda_h f_h(x)$. If $\lambda_j > 0$, $j=1,\ldots,p$, there exists a vector $\hat{\epsilon}$ such that $\hat{x}$ is an optimal solution of $\min\limits_{x\in \mathcal{X}} f_j(x)$, subject to $f_h(x)\leq \hat{\epsilon}_h, h=1,\ldots,p$, $h\ne j$, too.
\item[2.] Suppose that $\mathcal{X}$ is a convex set and $f_h:R^n \rightarrow R$ are convex functions. If $\hat{x}$ is an optimal solution of $\min\limits_{x\in \mathcal{X}} f_j(x)$, subject to $f_h(x)\leq \epsilon_h, h=1,\ldots,p$, $h\ne j$ for some $j$, there exists $\hat{\lambda}\in R^p_{\geq}$ such that $\hat{x}$ is optimal for $\min\limits_{x\in \mathcal{X}} \sum\limits_{h=1}^p \hat{\lambda}_h f_h(x)$.
\end{itemize}
\end{thm}

\noindent In view of Theorem \ref{Theorem_1}, we apply the $\epsilon$-constraint approach to our original three-objective portfolio selection model \eqref{eq:Triobjective_lambda_scalarized} by considering $\sigma^2_P(x)$ as the objective function. We obtain the following mathematical program:

\begin{equation}\label{eq:Triobjective_epsilon_scalarized}
		\begin{array}{llll}
            \min\limits_{(x,v,u)} & \sigma_{P}^{2}(x) & \\[5pt]
            \mbox{s.t.} & & & \\[5pt]
            & \mu_{P}(x)\ge\bar \mu & \\[5pt]
            & ku+\displaystyle\sum_{i=1}^{m}v_{i}\le \bar \gamma & \\[5pt]
            & v_{i}+u\ge (s^{i})^{T}x & i=1,\dots,m \\[5pt]
            & u \geq 0 & & \\[5pt]
            & v_{i}\ge 0 & i=1,\dots,m \\[5pt]
            & x\in\Omega & &
            \end{array}
\end{equation}
where $\bar\mu$ and $\bar\gamma$ are the required target levels for the portfolio expected return and $k-$Worst Non-ESG score, respectively.
Model \eqref{eq:Triobjective_epsilon_scalarized} is still a convex quadratic program that can be efficiently solved by applying standard optimization solvers (Cplex, Gurobi, etc.).
In the following section, we provide details on the methodology to obtain efficient portfolios of model \eqref{eq:Triobjective} by varying the targets $\bar\mu$ and $\bar\gamma$ in model (\ref{eq:Triobjective_epsilon_scalarized}): this produces an empirical construction of the efficient surface for \eqref{eq:Triobjective}. We then select several of such efficient portfolios to provide an empirical out-of-sample performance analysis.

%%%%%%%%%%%%%%%%%%%%%%%%%%%%%%%%%%%%%%%%%%%%%%%%%%%%%%%%%%%%%%%%%%%%%%%%%%%%%%%%%%%%%%%%%%%%%%%%%%%%%%%%%%%%%%%%%%%%%%%%%%%%%%%%
\subsection{Finding the efficient surface \label{subsec:EffSurface}}

Since in our multicriteria analysis we are interested in studying the relationship between the three considered criteria, we need to generate (approximations) of the set of efficient portfolios.

\noindent Consider a multicriteria minimization model with $p$ objectives and a feasible set $\mathcal{X}$ in the minimization form, that is, $\min\{f_1(x),\ldots,f_p(x)\; | x\in \mathcal{X}\}$. Formally, a feasible solution $\hat{x}\in \mathcal{X}$ is \emph{efficient} (or \emph{Pareto-optimal}) w.r.t. the $p$-objective model if there is no $x\in \mathcal{X}$ such that $f_h(x)\leq f_h(\hat{x})$, for $h=1,\ldots,p$, and $f_i(x)<f_i(\hat{x})$ for some $i\in {1,\ldots,p}$. We are interested in providing (an approximation of) the set of efficient solutions, which can be obtained by means either of a scalar weighted sum of the three objectives (i.e., as in model \eqref{eq:Triobjective_lambda_scalarized}), or through an $\epsilon$-constraint approach, that is, by optimizing one of the three objectives while moving the other two as constraints \citep[i.e., as in Model \eqref{eq:Triobjective_epsilon_scalarized}, see][]{ehrgott2005multicriteria}.

\noindent
In view of Theorem \ref{Theorem_1}, to construct the efficient frontier of our three-objective portfolio model \eqref{eq:Triobjective}, we can apply the $\epsilon$-constraint method by minimizing the portfolio variance in model (\ref{eq:Triobjective_epsilon_scalarized}) for different pairs of values fixed for the other two criteria.
For each fixed target value $\bar \mu$ for the portfolio expected return, we find the best compromise solution w.r.t. the portfolio variance and the Non-ESG score (with expected return at least equal to $\bar\mu$) and we represent the set of the efficient portfolios found as points in the plane $(\sigma^2, \gamma)$.
By changing the target value $\bar \mu$, we obtain several curves of this type  and provide the approximation of the efficient frontier of the three-objective model that, in fact, is as a surface in the three-dimension space $(\sigma^2, \gamma, \mu)$ \citep[see also][]{steuer2023non}.

\noindent
For $\bar \mu$, we first find the minimum variance portfolio in the standard Markowitz framework, and denote its expected return value by $\mu_{minV}$.
We also find the portfolio corresponding to the optimal solution of the following model:
\begin{equation}\label{min-score}
		\begin{array}{llll}
            \min\limits_{(x,v,u)} & ku+\displaystyle\sum_{i=1}^{m}v_{i} & \\[5pt]
            \mbox{s.t.} & & & \\[5pt]
            & v_{i}+u\ge (s^{i})^{T}x & i=1,\dots,m \\[5pt]
            & u \geq 0 & & \\[5pt]
            & v_{i}\ge 0 & i=1,\dots,m \\[5pt]
            & x\in\Omega & &		
            \end{array}
\end{equation}

\noindent
that is, the minimum Non-ESG score portfolio, whose expected return is denoted by $\mu_{minScore}$.
Then, we fix the minimum value of $\bar \mu$ as $\mu_{min}=\max \{\mu_{minV},\mu_{minScore}\}$.
On the other hand, the maximum possible value for $\bar \mu$, $\mu_{max}$, corresponds to the portfolio with the maximum expected return.
Therefore, to construct the efficient frontier, we fix a set of values for $\bar \mu$ in the interval $[\mu_{min}, \mu_{max}]$.
To obtain a surface, for each fixed level $\bar \mu \in [\mu_{min}, \mu_{max}]$, we define the appropriate interval of values for $\bar \gamma$ that we denote by $[\gamma_{min}(\bar \mu), \gamma_{max}(\bar \mu)]$.
The value $\gamma_{min}(\bar \mu)$ corresponds to the Non-ESG score of the optimal solution of the following model:

\begin{equation}\label{min-score-mu-constr}
		\begin{array}{llll}
            \min\limits_{(x,v,u)} & ku+\displaystyle\sum_{i=1}^{m}v_{i} & \\[5pt]
            \mbox{s.t.} & & & \\[5pt]
            & \mu_{P}(x)\ge\bar \mu & \\[5pt]
            & v_{i}+u\ge (s^{i})^{T}x & i=1,\dots,m \\[5pt]
            & u \geq 0 & & \\[5pt]
            & v_{i}\ge 0 & i=1,\dots,m \\[5pt]
            & x\in\Omega & &	
            \end{array}
\end{equation}

\noindent
while the value $\gamma_{max}(\bar \mu)$ is the Non-ESG score of the optimal solution of the following model:
\begin{equation}\label{min-var-muconstr}
		\begin{array}{llll}
            \min\limits_{(x,v,u)} & \sigma_{P}^{2}(x) & \\[5pt]
            \mbox{s.t.} & & & \\[5pt]
            & \mu_{P}(x)\ge\bar \mu & \\[5pt]
            & x\in\Omega & &
            \end{array}
\end{equation}

\begin{figure}[htbp!]
        \centering
	\includegraphics[width=0.8\linewidth]{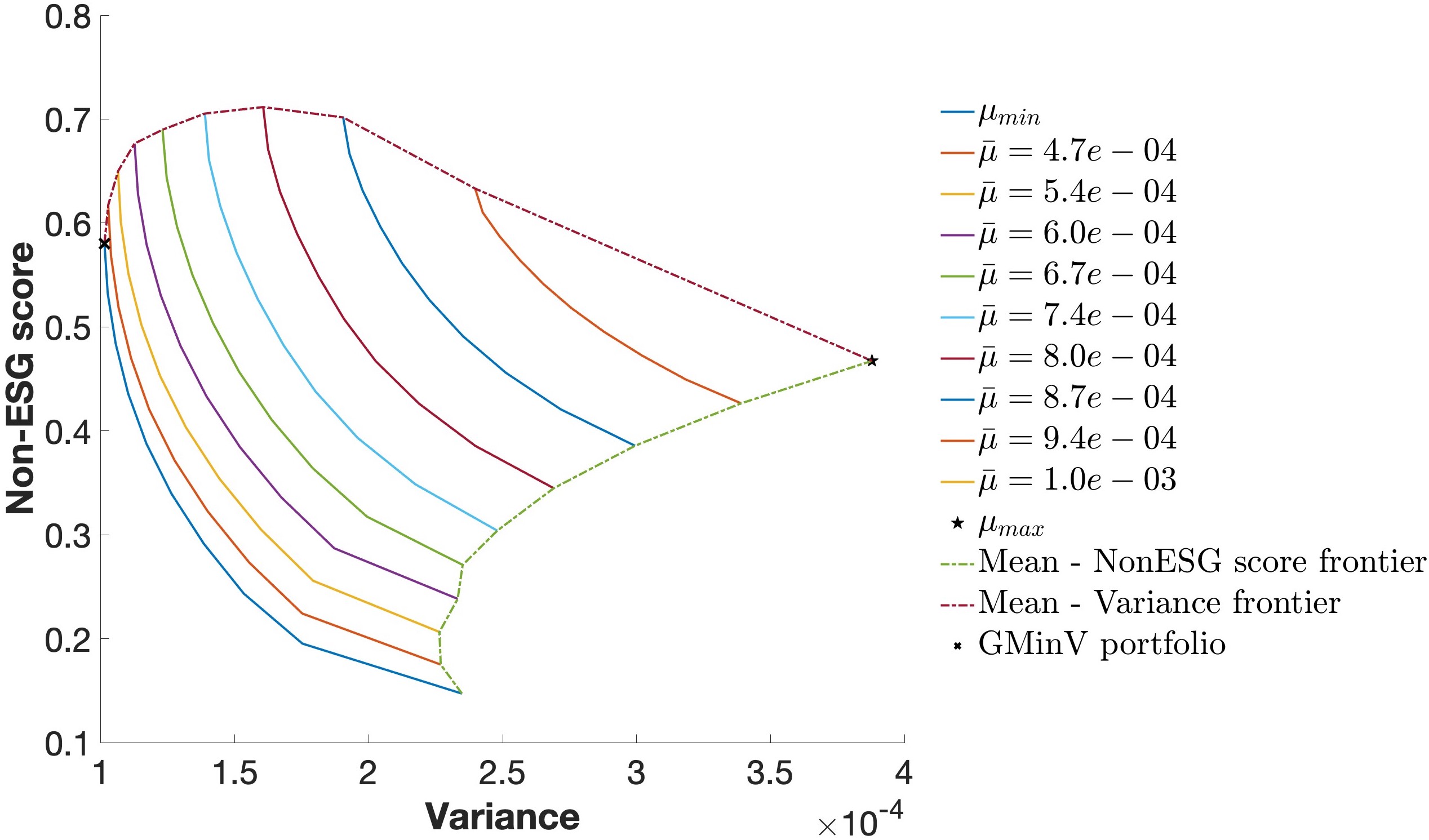}
	\caption{Example of the \emph{Mean-Variance-NonESG} efficient portfolios for several levels of the portfolio expected return $\bar\mu$ in the Variance\,-\,Non ESG score plane.}
 \label{fig:MVNonESG_colour}
\end{figure}

\noindent
In Figure \ref{fig:MVNonESG_colour} we report an example of the surface of the efficient portfolios obtained in the $(\sigma^2, \gamma)$ plane for several fixed levels of the expected return target $\bar\mu$, for the EuroStoxx50 dataset.

\noindent
Note that by solving model \eqref{eq:Triobjective_epsilon_scalarized} for different levels of the portfolio expected return $\bar\mu \in [\mu_{\min},\,\mu_{\max}]$ and with the corresponding $\bar\gamma=\gamma_{\min}(\bar\mu)$, we obtain efficient frontier in the $(\mu, \gamma)$ plane in Figure \ref{fig:MVNonESG_colour} (see the bold green dashed line). On the other hand, when we solve model \eqref{eq:Triobjective_epsilon_scalarized} for different values of $\bar\mu \in [\mu_{\min},\,\mu_{\max}]$, but with $\bar\gamma=\gamma_{\max}(\bar\mu)$, we obtain the Mean-Variance efficient frontier in the $(\mu, \gamma)$ plane (red dashed line in Figure \ref{fig:MVNonESG_colour}).

\noindent
For a fixed level of $\bar\mu \in [\mu_{\min},\,\mu_{\max}]$, if we require stronger conditions on the portfolio score,
namely lower levels of $\bar\gamma(\bar\mu)$,
we clearly obtain efficient portfolios with higher variance,
because the feasible region of \eqref{eq:Triobjective_epsilon_scalarized} becomes smaller.
Furthermore, as the required target portfolio return $\bar\mu$ increases, both the portfolio variance and its score tend to increase, too, as shown in Figure \ref{fig:MVNonESG_colour}.
We also note that if we fix in \eqref{eq:Triobjective_epsilon_scalarized} $\bar\mu=\mu_{\min}$ and $\bar\gamma=\gamma_{\max}(\mu_{\min})$, the optimal solution is the Global Minimum Variance (GMinV) portfolio (see the bold cross in figures \ref{fig:MVNonESG_colour} and \ref{fig:MVNonESG_transp}).
On the other hand, when $\bar\mu=\mu_{\max}$, the efficient frontier degenerates into a single point corresponding to the portfolio composed by the single asset with the highest expected return (the star in Figure \ref{fig:MVNonESG_colour}).

%%%%%%%%%%%%%%%%%%%%%%%%%%%%%%%%%%%%%%%%%%%%%%%%%%%%%%%%%%%%%%%%%%%%%%%%%%%%%%%%%%%%%%%%%%%%%%%%%%%%%%%%%%%%%%%%%%%%%%%%%%%%%%%%
\section{Empirical analysis for the evaluation of portfolio performance}\label{sec:EmpAnalysis}

\begin{figure}[htbp!]
        \centering
	\includegraphics[width=0.65\linewidth]{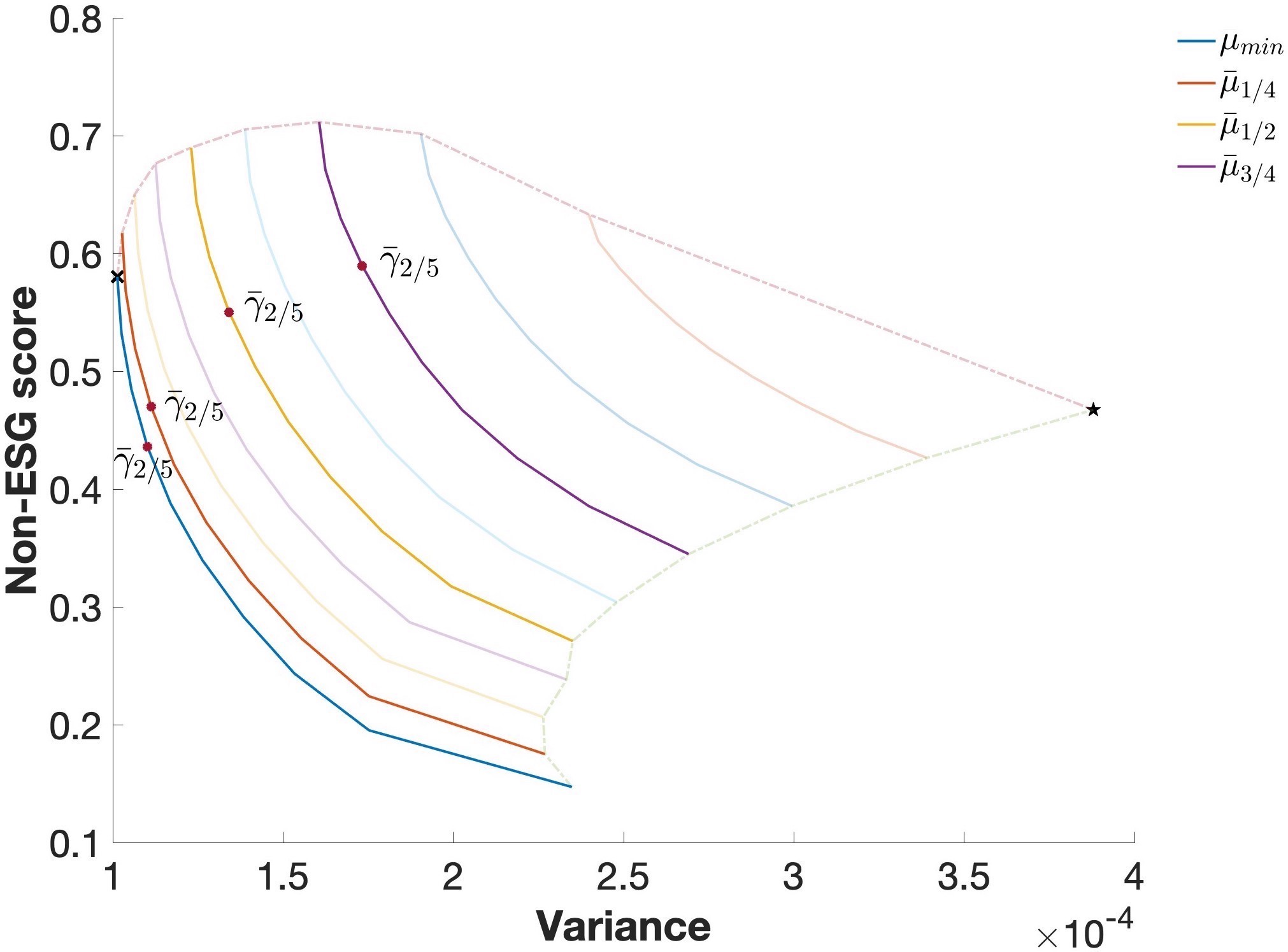}
	\caption{Example of the four \emph{Mean-Variance-NonESG} efficient portfolios selected for the out-of-sample performance analysis.}
 \label{fig:MVNonESG_transp}
\end{figure}

In this section we study the out-of sample performance of some of the efficient portfolios found in the previous section. Figure \ref{fig:MVNonESG_transp} shows the selected portfolios in the efficient surface of Figure \ref{fig:MVNonESG_colour}.

\medskip

\noindent
More precisely, our empirical analysis is based on four efficient portfolios which represent the optimal choice of agents with different profiles, according to their attitude toward risk, gain, and sustainability.
Indeed, the selected portfolios correspond to four different increasing levels of target return $\bar\mu_{\alpha}=\mu_{\min}+\alpha(\mu_{max}-\mu_{min})$, $\alpha=0,\frac{1}{4},\frac{1}{2},\frac{3}{4}$, while the target value for the portfolio sustainability is fixed at an intermediate value $\bar\gamma_{\frac{2}{5}}(\bar\mu_{\alpha})=\gamma_{min}(\bar\mu_{\alpha})+\frac{2}{5}(\gamma_{max}(\bar\mu_{\alpha})-\gamma_{min}(\bar\mu_{\alpha}))$.

We provide an extensive empirical analysis on real-world data sets. All the experiments have been performed on a workstation with Intel(R) Xeon(R) E5-2623 v4 CPU @ 2.60GHz processor and 64 GB of RAM, under MS Windows 10 Pro, using MATLAB R2022b and the GUROBI 9.5.1 optimization solver.

%%%%%%%%%%%%%%%%%%%%%%%%%%%%%%%%%%%%%%%%%%%%%%%%%%%%%%%%%%%%%%%%%%%%%%%%%%%%%%%%%%%%%%%%%%%%%%%%%%%%%%%%%%%%%%%%%%%%%%%%%%%%%%%%
\subsection{Description of the data sets and methodology \label{subsec:EmpSetup}}
The empirical analysis is based on a rolling time window (RTW) scheme of evaluation. We consider in-sample windows of 2 years (i.e., 500 observations), and we choose a one financial month period both as the rebalancing interval and the holding period.

\noindent
In Table \ref{tab:RatingProviders} we report the Rating Providers considered in this analysis, while in Table \ref{tab:DailyDatasets} we describe the data sets which consist of daily prices adjusted for dividends and stock splits for two major stock market indexes, i.e. EuroStoxx50 and NASDAQ100.

%The data sets have been supplied by Banco Mediolanum S.A.:
%
\begin{table}[htbp!]
	\small
	\centering
	%\begin{center}
	\scalebox{1.}{\begin{tabular}{l l l l l l l l l l l l l l l l l l}
			\toprule
			Rating Provider & & & & &  Time & & & & Frequency\\
			\midrule
			\text{Refinitiv} & & & & & \text{Jan} 2016--\text{Dec} 2021 & & & & \text{Monthly}\\
            \text{Bloomberg} & & & & & \text{Jan} 2016--\text{Dec} 2021 & & & & \text{Yearly}\\
            \text{S\&P Global} & & & & & \text{Jan} 2016--\text{Dec} 2021 & & & & \text{Yearly}\\
			\text{Morningstar Sustainalytics} & & & & & \text{Jan} 2016--\text{Dec} 2021 & & & & \text{Monthly} \\
			\bottomrule
	\end{tabular}}
	%\end{center}
	\caption{List of Rating Providers}
	\label{tab:RatingProviders}
\end{table}
\begin{table}[htbp!]
	\small
	\centering
	%\begin{center}
	\scalebox{1.}{\begin{tabular}{l l l l l l l l l l l l l l l l l l}
			\toprule
			Market Index & & & & &  $\#$Assets & & & & Country & & & & Time Interval\\
			\midrule
			\text{EuroStoxx50} & & & & & 46 & & & & \text{EU} & & & & \text{Jan} 2016--\text{Dec} 2021\\
			\text{NASDAQ100} & & & & & 70 & & & & \text{USA} & & & & \text{Jan} 2016--\text{Dec} 2021 \\
			\bottomrule
	\end{tabular}}
	%\end{center}
	\caption{List of data sets}
	\label{tab:DailyDatasets}
\end{table}

\noindent
For comparison purposes, in our empirical analysisi we also consider other standard portfolio selection strategies. In Table \ref{tab:ListOfModels} we summarize all the analyzed methods. In particular, we compare with a previous optimization approach presented in \cite{cesarone2022does}, in which a single ESG evaluation was considered (only scores given by Refinitiv) in the classical Markowitz portfolio selection framework by including a target constraint on the ESG evaluation of the portfolio, as well as, on the portfolio expected return. We refer to this model as the Mean-Variance-ESG optimization model (MV-ESG). On the other hand, we refer to our model (\ref{eq:Triobjective_epsilon_scalarized}) as Mean Variance $k-$Worst NonESG (MV $k-$Worst NonESG). For both models, in our notation the subscript from 1 to 4 indicates the increasing target level for the portfolio expected return.

\begin{table}[htbp]
	\centering
	\scalebox{0.85}{
	\begin{tabular}{llll}
		\toprule
		Approach & & Abbreviation \\
		\midrule
		Global Minimum-Variance portfolio & & GMinV \\
                 Equally Weighted & & EW \\
                 Risk Parity & & RP \\
                 Most Diversified Portfolio & & MDP \\
                  & & \\
                  & & \\
                  \textit{\textbf{Mean-Variance-ESG optimization strategy (Cesarone et al. 2022)}} \\
                  MV-ESG: Low Gain ($\eta_{0}$) & & Sust$\_{1}$ \\
		MV-ESG: Intermediate Low Gain ($\eta_{1/4}$) & & Sust$\_{2}$ \\
		MV-ESG: Intermediate High Gain ($\eta_{1/2}$) & & Sust$\_{3}$ \\
		MV-ESG: High Gain ($\eta_{3/4}$) & & Sust$\_{4}$\\[5pt]
		\textit{$\boldsymbol{k-sum}$ $\textbf{optimization strategy}$} \\
		MV $k-$Worst NonESG: Low Gain ($\bar{\mu}_{0}$) & & Sust$\_{1\_k}$Worst \\
		MV $k-$Worst NonESG: Intermediate Low Gain ($\bar{\mu}_{1/4}$) & & Sust$\_{2\_k}$Worst \\
		MV $k-$Worst NonESG: Intermediate High Gain ($\bar{\mu}_{1/2}$) & & Sust$\_{3\_k}$Worst \\
		MV $k-$Worst NonESG: High Gain ($\bar{\mu}_{3/4}$) & & Sust$\_{4\_k}$Worst
	\end{tabular}%
    }
    \caption{Summary of portfolio selection strategies}
	\label{tab:ListOfModels}%
\end{table}

\noindent
The out-of-sample performance results of each portfolio strategy are evaluated by considering several performance measures widely used in the literature \citep[see, e.g.,][]{cesarone2022comparing,cesarone2022does,cesarone2023mean}, and detailed below.\\
\vskip 8 pt
\noindent The Sharpe ratio (\textbf{Sharpe}, \citep{sharpe1966mutual,Sharpe1994}) measures the gain per unit of risk and is defined as follows:
\begin{equation*}
\mbox{Sharpe} = \frac{\mu^{out}-r_{f}}{\sigma^{out}},
\end{equation*}
where $r_{f}=0$, $\mu^{out}$ (\textbf{ExpRet}) is the expected value of the random variable portfolio returns $R^{out}$ computed over the out-of-sample period, and $\sigma^{out}$ (\textbf{Vol}) is its standard deviation. Clearly, higher Sharpe ratios indicate better portfolio performances.
\vskip 8 pt
\noindent The Maximum DrawDown \citep[\textbf{MDD},][]{Chekhlov2005} measures the maximum potential out-of-sample loss from the observed peak, and is defined as
\begin{equation*}
\mbox{MDD} = \min\limits_{T^{in}+1\le t\le T} DD_{t},
\end{equation*}
where $T^{in}$ is the length of the in-sample window, and, at each time $t$, the DrawDown $DD_t$ is computed as
\begin{equation*}
DD_{t}=\frac{W_{t}-\max\limits_{T^{in}+1\le \tau \le t}W_{\tau}}{\max\limits_{T^{in}+1\le \tau \le t}W_{\tau}}, \qquad t \in \{T^{in}+1,\dots T\},
\end{equation*}
where $W_{0}=1$ and $W_{t}=W_{t-1}(1+R^{out}_{t})$ denotes the portfolio wealth at time $t$. The MDD is always non-positive, hence values close to $0$ are preferable.
\vskip 8 pt
\noindent The Ulcer index (\textbf{Ulcer}, \citep{martin1989investor}) evaluates the depth and the duration of DrawDowns over the out-of-sample period and is defined as
\begin{equation*}
\mbox{Ulcer} = \sqrt{\frac{\sum\limits_{t=T^{in}+1}^{T}DD_{t}^2}{T-T^{in}}}.
\end{equation*}
A lower Ulcer value suggests a better portfolio performance.

\noindent The Rachev ratio \citep[\textbf{Rachev10},][]{RacBigOrtSto2004} measures the relative gap between the mean of the best $\alpha\%$ values of $R^{out}-r_{f}$ and that of the worst $\beta\%$ ones, and it is computed as
\begin{equation*}
\mbox{Rachev10} = \frac{CVaR_{\alpha}(r_{f}-R^{out})}{CVaR_{\beta}(R^{out}-r_{f})},
\end{equation*}
where $\alpha=\beta=10\%$, and $r_{f} = 0$. Higher Rachev ratio values are clearly preferred.
\vskip 8 pt
\noindent The Turnover \citep[\textbf{Turn},][]{Demiguel2009} estimates the amount of trading required to follow the portfolio optimization strategy along the successive time windows and to rebalance the portfolio fractions, accordingly. It is defined as follows
\begin{equation*}
\mbox{Turn} = \frac{1}{Q}\sum_{q=1}^{Q}\sum_{j=1}^{n}\mid x_{q,j}-x_{q-1,j}\mid ,
\end{equation*}
where $Q$ is the number of rebalances, $x_{q,j}$ is the portfolio weight of asset $j$ after rebalancing, and $x_{q-1,j}$ is the portfolio weight before rebalancing at time $q$. Lower turnover values indicate better portfolio performance. We point out that this definition of portfolio turnover is a proxy of the effective one, since it evaluates only the amount of trading generated by the models at each rebalance, without considering the trades due to changes in asset
prices between one rebalance and the next. Thus, by definition, the turnover of the EW portfolio is zero.
\vskip 8 pt
\noindent The Jensen's Alpha \citep[\textbf{AlphaJ},][]{jensen1968performance} is defined as the intercept of the line given by the linear regression of the portfolio return $R^{out}$ on the expected value of the market index return $R_{I}^{out}$, namely

\begin{equation*}
\mbox{AlphaJ} = \mathbb{E}[R^{out}]-\beta\mathbb{E}[R_{I}^{out}]\,,
\end{equation*}
where $\displaystyle\beta=\frac{Cov(R^{out},R_{I}^{out})}{\sigma^{2}(R_{I}^{out})}$.
\vskip 8 pt
\noindent The Information ratio \citep[\textbf{InfoR},][]{treynor1973use} is defined as the expected value of the random variable given by the difference between the out-of-sample portfolio return and that of the benchmark index, divided by the standard deviation of such difference, namely
\begin{equation*}
\mbox{InfoR} = \frac{\mathbb{E}[R^{out}-R_{I}^{out}]}{\sigma[R^{out}-R_{I}^{out}]}.
\end{equation*}
Clearly, the larger its value, the better the portfolio performance.

\noindent
The Value-at-Risk \citep[\textbf{VaR5},][]{longerstaey1996riskmetricstm} represents the maximum potential loss at a given confidence level $\varepsilon$ over the out-of-sample time period $L$, and is defined as
\begin{equation*}
\mbox{VaR5}=\mbox{max}^{\lfloor\varepsilon L\rfloor+1}\{-R_{T^{in}+1}^{out},\dots,-R_{T}^{out}\}\,,
\end{equation*}
where we set $\varepsilon=5\%$ and with $L=T-T^{in}$.
\vskip 8 pt
\noindent The Omega ratio \citep[\textbf{Omega},][]{harlow1989asset} is the ratio between the average of positive and negative out-of-sample portfolio returns, namely

\begin{equation*}
\mbox{Omega} = \frac{\mathbb{E}[\max\{0,R^{out}-\phi\}]}{\mathbb{E}[\min\{0,R^{out}-\phi\}]}\,,
\end{equation*}
with $\phi=0$ in our experiments.
\vskip 8 pt
\noindent The average number of selected assets (\textbf{ave$\#$}) is considered as and indicator of the level of portfolio diversification.
\vskip 8 pt
\noindent The Return on Investment (ROI) measures the time-by-time return generated by each portfolio strategy over a given time horizon $\Delta\tau$, and is defined as follows
\begin{equation*}
ROI_{P,\tau}=\frac{W_{P,\tau}-W_{P,\tau-\Delta\tau}}{W_{P,\tau-\Delta\tau}}\qquad \tau=\Delta\tau+1,\dots,T\,.
\end{equation*}
\noindent Here, $W_{P,\tau-\Delta\tau}$ denotes the amount of capital invested at the beginning of the investment horizon, $W_{P,\tau}=W_{P,\tau-\Delta\tau}\prod_{t=\tau-\Delta\tau+1}^{\tau}(1+R_{t}^{out})$ indicates the portfolio wealth, and $T$ is the number of historical scenarios.

%%%%%%%%%%%%%%%%%%%%%%%%%%%%%%%%%%%%%%%%%%%%%%%%%%%%%%%%%%%%%%%%%%%%%%%%%%%%%%%%%%%%%%%%%%%%%%%%%%%%%%%%%%%%%%%%%%%%%%%%%%%%%%%%
\subsection{Out-of-sample performance results \label{subsec:OutOfSample}}

In the following tables we provide the computational results obtained in the application of our model (\ref{eq:Triobjective_epsilon_scalarized}) to the EuroStoxx50 and NASDAQ100 data sets, as well as, those related to all the other portfolio selection strategies listed in Table \ref{tab:ListOfModels}.
We report the case where $k=1$ in our $k-$sum optimization strategy; this corresponds to minimize the worst portfolio score among those assigned by the four considered agencies.
For the sake of readability, the cases with $k=2,3,4$, are reported in the Appendix as supplementary materials.
\vskip 7 pt
We observe that choosing $k=1$ does not correspond to considering the score of a single agency for all assets, but, according to its set of ratings and the corresponding ordering, the worst score of each asset may derive from different agencies' evaluations.

\begin{table}[htbp]
  \centering
  \caption{EuroStoxx50 (with $k = 1$)}
    \scalebox{0.7}{\hskip-1.65cm\begin{tabular}{|c|c|c|c|c|c|c|c|c|c|c|c|c|}
    \toprule
    \textbf{Approach} & \textbf{ExpRet} & \textbf{Vol} & \textbf{Sharpe} & \textbf{MDD} & \textbf{Ulcer} & \textbf{Rachev10} & \textbf{Turn} & \textbf{AlphaJ} & \textbf{InfoRatio} & \textbf{VaR5} & \textbf{Omega} & \textbf{ave \#} \\
    \midrule
    \textbf{MV0} & \cellcolor[rgb]{ .984,  .694,  .471}0.042\% & \cellcolor[rgb]{ .388,  .745,  .482}0.939\% & \cellcolor[rgb]{ .992,  .812,  .494}4.45\% & \cellcolor[rgb]{ .996,  .875,  .506}-0.321 & \cellcolor[rgb]{ .996,  .851,  .506}8.53\% & \cellcolor[rgb]{ .996,  .898,  .51}0.920 & \cellcolor[rgb]{ .671,  .824,  .498}0.13 & \cellcolor[rgb]{ .988,  .733,  .478}0.025\% & \cellcolor[rgb]{ .98,  .62,  .459}2.46\% & \cellcolor[rgb]{ .388,  .745,  .482}1.22\% & \cellcolor[rgb]{ .992,  .839,  .498}1.145 & 12 \\
    \midrule
    \textbf{EW} & \cellcolor[rgb]{ .984,  .663,  .467}0.040\% & \cellcolor[rgb]{ .98,  .541,  .447}1.297\% & \cellcolor[rgb]{ .973,  .459,  .427}3.11\% & \cellcolor[rgb]{ .973,  .412,  .42}-0.407 & \cellcolor[rgb]{ .984,  .58,  .455}9.44\% & \cellcolor[rgb]{ .996,  .918,  .514}0.919 & -     & \cellcolor[rgb]{ .973,  .471,  .431}0.014\% & \cellcolor[rgb]{ .537,  .788,  .494}7.29\% & \cellcolor[rgb]{ .992,  .722,  .482}1.78\% & \cellcolor[rgb]{ .976,  .502,  .435}1.105 & 44 \\
    \midrule
    \textbf{RP} & \cellcolor[rgb]{ .984,  .639,  .463}0.039\% & \cellcolor[rgb]{ .992,  .706,  .478}1.201\% & \cellcolor[rgb]{ .976,  .498,  .435}3.26\% & \cellcolor[rgb]{ .976,  .518,  .439}-0.387 & \cellcolor[rgb]{ .992,  .773,  .49}8.79\% & \cellcolor[rgb]{ .988,  .714,  .475}0.905 & \cellcolor[rgb]{ .388,  .745,  .482}0.02 & \cellcolor[rgb]{ .976,  .49,  .431}0.015\% & \cellcolor[rgb]{ .427,  .757,  .486}7.77\% & \cellcolor[rgb]{ 1,  .871,  .51}1.60\% & \cellcolor[rgb]{ .976,  .545,  .443}1.110 & 44 \\
    \midrule
    \textbf{MDP} & \cellcolor[rgb]{ .996,  .875,  .506}0.050\% & \cellcolor[rgb]{ .843,  .875,  .506}1.040\% & \cellcolor[rgb]{ .996,  .91,  .514}4.82\% & \cellcolor[rgb]{ .992,  .816,  .494}-0.331 & \cellcolor[rgb]{ .424,  .753,  .482}6.08\% & \cellcolor[rgb]{ 1,  .922,  .518}0.920 & \cellcolor[rgb]{ .671,  .824,  .498}0.13 & \cellcolor[rgb]{ .996,  .867,  .506}0.031\% & \cellcolor[rgb]{ .992,  .808,  .494}4.26\% & \cellcolor[rgb]{ .875,  .882,  .51}1.47\% & \cellcolor[rgb]{ 1,  .922,  .518}1.155 & 17 \\
    \midrule
    \textbf{Sust\_1} & \cellcolor[rgb]{ .973,  .412,  .42}0.029\% & \cellcolor[rgb]{ .553,  .792,  .49}0.976\% & \cellcolor[rgb]{ .973,  .412,  .42}2.93\% & \cellcolor[rgb]{ .996,  .851,  .502}-0.324 & \cellcolor[rgb]{ .973,  .412,  .42}10\% & \cellcolor[rgb]{ .984,  .631,  .459}0.899 & \cellcolor[rgb]{ .765,  .851,  .502}0.17 & \cellcolor[rgb]{ .973,  .412,  .42}0.011\% & \cellcolor[rgb]{ .973,  .412,  .42}0.50\% & \cellcolor[rgb]{ .655,  .82,  .494}1.36\% & \cellcolor[rgb]{ .973,  .412,  .42}1.094 & 16 \\
    \midrule
    \textbf{Sust\_2} & \cellcolor[rgb]{ .976,  .553,  .447}0.035\% & \cellcolor[rgb]{ .569,  .796,  .49}0.979\% & \cellcolor[rgb]{ .98,  .584,  .451}3.60\% & \cellcolor[rgb]{ .996,  .878,  .506}-0.320 & \cellcolor[rgb]{ .984,  .6,  .459}9.36\% & \cellcolor[rgb]{ .973,  .412,  .42}0.883 & \cellcolor[rgb]{ .867,  .882,  .51}0.22 & \cellcolor[rgb]{ .98,  .561,  .447}0.018\% & \cellcolor[rgb]{ .976,  .518,  .439}1.53\% & \cellcolor[rgb]{ .698,  .835,  .498}1.38\% & \cellcolor[rgb]{ .98,  .588,  .451}1.115 & 15 \\
    \midrule
    \textbf{Sust\_3} & \cellcolor[rgb]{ .969,  .914,  .518}0.054\% & \cellcolor[rgb]{ 1,  .867,  .51}1.108\% & \cellcolor[rgb]{ .992,  .922,  .518}4.90\% & \cellcolor[rgb]{ .596,  .808,  .498}-0.301 & \cellcolor[rgb]{ .847,  .875,  .506}7.70\% & \cellcolor[rgb]{ .98,  .588,  .451}0.896 & \cellcolor[rgb]{ .988,  .671,  .471}0.36 & \cellcolor[rgb]{ .965,  .914,  .518}0.036\% & \cellcolor[rgb]{ .992,  .784,  .49}4.03\% & \cellcolor[rgb]{ .992,  .737,  .482}1.76\% & \cellcolor[rgb]{ .996,  .918,  .514}1.154 & 11 \\
    \midrule
    \textbf{Sust\_4} & \cellcolor[rgb]{ .733,  .847,  .506}0.070\% & \cellcolor[rgb]{ .98,  .506,  .439}1.318\% & \cellcolor[rgb]{ .906,  .894,  .514}5.28\% & \cellcolor[rgb]{ .706,  .839,  .502}-0.304 & \cellcolor[rgb]{ 1,  .886,  .514}8.41\% & \cellcolor[rgb]{ .961,  .91,  .518}0.922 & \cellcolor[rgb]{ .976,  .482,  .435}0.43 & \cellcolor[rgb]{ .753,  .851,  .506}0.048\% & \cellcolor[rgb]{ .996,  .91,  .514}5.21\% & \cellcolor[rgb]{ .976,  .439,  .427}2.13\% & \cellcolor[rgb]{ .949,  .91,  .518}1.162 & 7 \\
    \midrule
    \textbf{Sust$\_{\boldsymbol{1}\_k}$}\textbf{Worst} & \cellcolor[rgb]{ .831,  .875,  .51}0.063\% & \cellcolor[rgb]{ .537,  .788,  .49}0.972\% & \cellcolor[rgb]{ .62,  .812,  .498}6.52\% & \cellcolor[rgb]{ .624,  .816,  .498}-0.302 & \cellcolor[rgb]{ .4,  .745,  .482}5.98\% & \cellcolor[rgb]{ .388,  .745,  .482}0.956 & \cellcolor[rgb]{ 1,  .922,  .518}0.27 & \cellcolor[rgb]{ .776,  .859,  .506}0.047\% & \cellcolor[rgb]{ .98,  .918,  .518}5.38\% & \cellcolor[rgb]{ .471,  .769,  .486}1.26\% & \cellcolor[rgb]{ .565,  .796,  .494}1.217 & 12 \\
    \midrule
    \textbf{Sust$\_{\boldsymbol{2}\_k}$}\textbf{Worst} & \cellcolor[rgb]{ .741,  .847,  .506}0.069\% & \cellcolor[rgb]{ .624,  .812,  .494}0.992\% & \cellcolor[rgb]{ .518,  .784,  .49}6.97\% & \cellcolor[rgb]{ .463,  .769,  .49}-0.298 & \cellcolor[rgb]{ .388,  .745,  .482}5.93\% & \cellcolor[rgb]{ .573,  .8,  .494}0.945 & \cellcolor[rgb]{ 1,  .871,  .51}0.29 & \cellcolor[rgb]{ .69,  .831,  .502}0.052\% & \cellcolor[rgb]{ .8,  .867,  .51}6.17\% & \cellcolor[rgb]{ .604,  .804,  .494}1.33\% & \cellcolor[rgb]{ .459,  .769,  .49}1.233 & 12 \\
    \midrule
    \textbf{Sust$\_{\boldsymbol{3}\_k}$}\textbf{Worst} & \cellcolor[rgb]{ .471,  .769,  .49}0.087\% & \cellcolor[rgb]{ .996,  .784,  .494}1.157\% & \cellcolor[rgb]{ .388,  .745,  .482}7.51\% & \cellcolor[rgb]{ .388,  .745,  .482}-0.296 & \cellcolor[rgb]{ .576,  .796,  .49}6.66\% & \cellcolor[rgb]{ .831,  .875,  .51}0.930 & \cellcolor[rgb]{ .984,  .627,  .463}0.37 & \cellcolor[rgb]{ .427,  .757,  .486}0.068\% & \cellcolor[rgb]{ .388,  .745,  .482}7.93\% & \cellcolor[rgb]{ .988,  .675,  .471}1.84\% & \cellcolor[rgb]{ .388,  .745,  .482}1.243 & 8 \\
    \midrule
    \textbf{Sust$\_{\boldsymbol{4}\_k}$}\textbf{Worst} & \cellcolor[rgb]{ .388,  .745,  .482}0.092\% & \cellcolor[rgb]{ .973,  .412,  .42}1.373\% & \cellcolor[rgb]{ .576,  .8,  .494}6.70\% & \cellcolor[rgb]{ .698,  .835,  .502}-0.304 & \cellcolor[rgb]{ .965,  .91,  .514}8.14\% & \cellcolor[rgb]{ .914,  .898,  .514}0.925 & \cellcolor[rgb]{ .973,  .412,  .42}0.45 & \cellcolor[rgb]{ .388,  .745,  .482}0.070\% & \cellcolor[rgb]{ .498,  .78,  .49}7.46\% & \cellcolor[rgb]{ .973,  .412,  .42}2.16\% & \cellcolor[rgb]{ .627,  .816,  .498}1.209 & 6 \\
    \bottomrule
    \end{tabular}}%
  \label{tab:EuroStoxx50_Perf_q1}%
\end{table}%

\noindent
Table \ref{tab:EuroStoxx50_Perf_q1} shows the performance results of the selected portfolio strategies, obtained with the EuroStoxx50 data set.
Different colors are used to emphasize the good (green) and bad (red) performance results. For each performance index, the color spans from deep-green, representing the best result, to deep-red, representing the worst one. We observe that including a target on the portfolio sustainability tends to improve the overall performance with respect to the classical portfolio strategies, in terms of almost all indexes.
Among the two approaches which consider the portfolio sustainability, our $k-$Worst NonESG portfolios globally provide the best results.
A dark red result is registered only for the case with the higher target value for the expected return. Under this strict requirement, an high value also for portfolio volatility is a natural effect of constrained optimization models which, to guarantee the required return, necessarily have to choose an high volatility portfolio. This is clearly not the case of models MV0, EW, RP and MDP which do not include any condition on the expected portfolio return. Also for the Turnover index, it is typical that, to get the target, all Mean-Variance-ESG models - in fact, both ours and those by \cite{cesarone2022does} - produce higher values than the other methods. The only weakness of our models is related to the maximum out-of-sample loss measured by index VaR5 which for ``Sust$\_{{4}\_k}${Worst}'' is the highest. This is a systematic result obtained for this model with all possible values of $k$ and on both the two analyzed data sets (see the Appendix). In spite of this, this result is always registered for optimization models with a high value for the target on the expected return. This suggests that, as before, this is intrinsically related to constrained optimization.

\noindent The good performance of our models is confirmed by the out-of-sample value of ROI based on a 3-year time horizon that we report in Table \ref{tab:EuroStoxx50_ROI_q1}.

\begin{table}[htbp!]
  \centering
  \caption{ROI based on a 3-year time horizon (with $k = 1$)}
\scalebox{0.7}{\begin{tabular}{|c|c|c|c|c|c|c|c|}
    \toprule
    \textbf{Approach} & \textbf{ExpRet} & \textbf{Vol} & \textbf{5\%-perc} & \textbf{25\%-perc} & \textbf{50\%-perc} & \textbf{75\%-perc} & \textbf{95\%-perc} \\
    \midrule
    \textbf{MV0} & \cellcolor[rgb]{ .98,  .588,  .451}21\% & \cellcolor[rgb]{ .741,  .843,  .502}11\% & \cellcolor[rgb]{ .98,  .612,  .455}9\% & \cellcolor[rgb]{ .976,  .533,  .443}13\% & \cellcolor[rgb]{ .976,  .541,  .443}17\% & \cellcolor[rgb]{ .98,  .624,  .459}30\% & \cellcolor[rgb]{ .984,  .647,  .463}41\% \\
    \midrule
    \textbf{EW} & \cellcolor[rgb]{ .988,  .761,  .486}30\% & \cellcolor[rgb]{ 1,  .871,  .51}16\% & \cellcolor[rgb]{ .973,  .412,  .42}1\% & \cellcolor[rgb]{ .988,  .714,  .475}19\% & \cellcolor[rgb]{ .996,  .882,  .51}31\% & \cellcolor[rgb]{ .992,  .824,  .498}43\% & \cellcolor[rgb]{ .992,  .839,  .502}56\% \\
    \midrule
    \textbf{RP} & \cellcolor[rgb]{ .988,  .725,  .478}28\% & \cellcolor[rgb]{ .973,  .914,  .514}14\% & \cellcolor[rgb]{ .976,  .486,  .431}4\% & \cellcolor[rgb]{ .984,  .694,  .475}19\% & \cellcolor[rgb]{ .992,  .804,  .494}28\% & \cellcolor[rgb]{ .988,  .773,  .486}39\% & \cellcolor[rgb]{ .992,  .78,  .49}51\% \\
    \midrule
    \textbf{MDP} & \cellcolor[rgb]{ .996,  .918,  .514}38\% & \cellcolor[rgb]{ .812,  .867,  .506}12\% & \cellcolor[rgb]{ .953,  .91,  .518}25\% & \cellcolor[rgb]{ .973,  .914,  .518}28\% & \cellcolor[rgb]{ .996,  .914,  .514}33\% & \cellcolor[rgb]{ .996,  .902,  .514}48\% & \cellcolor[rgb]{ .996,  .878,  .51}59\% \\
    \midrule
    \textbf{Sust\_1} & \cellcolor[rgb]{ .973,  .412,  .42}13\% & \cellcolor[rgb]{ .388,  .745,  .482}5\% & \cellcolor[rgb]{ .976,  .498,  .435}4\% & \cellcolor[rgb]{ .973,  .412,  .42}9\% & \cellcolor[rgb]{ .973,  .412,  .42}12\% & \cellcolor[rgb]{ .973,  .412,  .42}16\% & \cellcolor[rgb]{ .973,  .412,  .42}23\% \\
    \midrule
    \textbf{Sust\_2} & \cellcolor[rgb]{ .976,  .498,  .435}17\% & \cellcolor[rgb]{ .482,  .773,  .486}7\% & \cellcolor[rgb]{ .98,  .565,  .447}7\% & \cellcolor[rgb]{ .976,  .494,  .435}12\% & \cellcolor[rgb]{ .976,  .506,  .435}16\% & \cellcolor[rgb]{ .976,  .494,  .435}21\% & \cellcolor[rgb]{ .976,  .494,  .435}29\% \\
    \midrule
    \textbf{Sust\_3} & \cellcolor[rgb]{ 1,  .922,  .518}38\% & \cellcolor[rgb]{ 1,  .91,  .518}15\% & \cellcolor[rgb]{ .996,  .871,  .506}21\% & \cellcolor[rgb]{ .996,  .878,  .51}25\% & \cellcolor[rgb]{ 1,  .922,  .518}33\% & \cellcolor[rgb]{ .988,  .922,  .518}50\% & \cellcolor[rgb]{ .973,  .914,  .518}65\% \\
    \midrule
    \textbf{Sust\_4} & \cellcolor[rgb]{ .804,  .867,  .51}51\% & \cellcolor[rgb]{ .988,  .69,  .475}20\% & \cellcolor[rgb]{ .863,  .882,  .51}29\% & \cellcolor[rgb]{ .875,  .886,  .514}33\% & \cellcolor[rgb]{ .827,  .875,  .51}43\% & \cellcolor[rgb]{ .78,  .859,  .506}68\% & \cellcolor[rgb]{ .769,  .855,  .506}88\% \\
    \midrule
    \textbf{Sust$\_{\boldsymbol{1}\_k}$}\textbf{Worst} & \cellcolor[rgb]{ .827,  .875,  .51}49\% & \cellcolor[rgb]{ .929,  .898,  .51}13\% & \cellcolor[rgb]{ .851,  .878,  .51}29\% & \cellcolor[rgb]{ .733,  .847,  .506}40\% & \cellcolor[rgb]{ .776,  .859,  .506}47\% & \cellcolor[rgb]{ .882,  .89,  .514}60\% & \cellcolor[rgb]{ .918,  .898,  .514}71\% \\
    \midrule
    \textbf{Sust$\_{\boldsymbol{2}\_k}$}\textbf{Worst} & \cellcolor[rgb]{ .745,  .851,  .506}55\% & \cellcolor[rgb]{ 1,  .894,  .514}15\% & \cellcolor[rgb]{ .749,  .851,  .506}34\% & \cellcolor[rgb]{ .671,  .827,  .502}43\% & \cellcolor[rgb]{ .698,  .835,  .502}52\% & \cellcolor[rgb]{ .808,  .867,  .51}66\% & \cellcolor[rgb]{ .82,  .871,  .51}82\% \\
    \midrule
    \textbf{Sust$\_{\boldsymbol{3}\_k}$}\textbf{Worst} & \cellcolor[rgb]{ .42,  .757,  .486}76\% & \cellcolor[rgb]{ .984,  .6,  .459}23\% & \cellcolor[rgb]{ .388,  .745,  .482}49\% & \cellcolor[rgb]{ .388,  .745,  .482}57\% & \cellcolor[rgb]{ .388,  .745,  .482}71\% & \cellcolor[rgb]{ .475,  .773,  .49}95\% & \cellcolor[rgb]{ .494,  .776,  .49}118\% \\
    \midrule
    \textbf{Sust$\_{\boldsymbol{4}\_k}$}\textbf{Worst} & \cellcolor[rgb]{ .388,  .745,  .482}78\% & \cellcolor[rgb]{ .973,  .412,  .42}28\% & \cellcolor[rgb]{ .467,  .769,  .49}46\% & \cellcolor[rgb]{ .431,  .757,  .486}55\% & \cellcolor[rgb]{ .392,  .749,  .486}70\% & \cellcolor[rgb]{ .388,  .745,  .482}102\% & \cellcolor[rgb]{ .388,  .745,  .482}129\% \\
    \bottomrule
    \end{tabular}}%
  \label{tab:EuroStoxx50_ROI_q1}%
\end{table}%

\noindent In Table \ref{tab:NASDAQ100_Perf_q1} we provide the computational results related to the NASDAQ100 dataset.
Similarly to the previous case, the better values for the indexes are obtained when portfolio sustainability is included in the model. However, for the NASDAQ100 data set we have less robust results since several dark red results are observed. This is probably due to the fact that, as we already discussed in Section \ref{sec:Disagreement}, in this market, there is some agreement among the stocks ESG evaluations by the different rating agencies (see Figure \ref{fig:ESG_Disagreement_NASDAQ100}). This also implies that the difference in the performance of the Mean-Variance-ESG approach by \cite{cesarone2022does} and our $k-$sum optimization strategy is less evident, even if the latter still performs better than the former.

% Table generated by Excel2LaTeX from sheet 'Foglio1'
\begin{table}[htbp!]
  \centering
  \caption{NASDAQ100 (with $k = 1$)}
    \scalebox{0.7}{\hskip-1.65cm\begin{tabular}{|c|c|c|c|c|c|c|c|c|c|c|c|c|}
    \toprule
    \textbf{Approach} & \textbf{ExpRet} & \textbf{Vol} & \textbf{Sharpe} & \textbf{MDD} & \textbf{Ulcer} & \textbf{Rachev10} & \textbf{Turn} & \textbf{AlphaJ} & \textbf{InfoRatio} & \textbf{VaR5} & \textbf{Omega} & \textbf{ave \#} \\
    \midrule
    \textbf{MV0} & \cellcolor[rgb]{ .973,  .42,  .42}0.045\% & \cellcolor[rgb]{ .388,  .745,  .482}1.124\% & \cellcolor[rgb]{ .973,  .424,  .42}3.99\% & \cellcolor[rgb]{ .996,  .906,  .514}-0.319 & \cellcolor[rgb]{ .902,  .89,  .51}6.83\% & \cellcolor[rgb]{ .973,  .478,  .431}0.901 & \cellcolor[rgb]{ .624,  .812,  .494}0.15 & \cellcolor[rgb]{ .973,  .439,  .424}-0.013\% & \cellcolor[rgb]{ .973,  .424,  .42}-5.80\% & \cellcolor[rgb]{ .388,  .745,  .482}1.43\% & \cellcolor[rgb]{ .973,  .435,  .424}1.145 & 19 \\
    \midrule
    \textbf{EW} & \cellcolor[rgb]{ .969,  .914,  .518}0.108\% & \cellcolor[rgb]{ 1,  .894,  .514}1.435\% & \cellcolor[rgb]{ .957,  .91,  .518}7.53\% & \cellcolor[rgb]{ .922,  .902,  .514}-0.312 & \cellcolor[rgb]{ .608,  .808,  .494}5.61\% & \cellcolor[rgb]{ .992,  .839,  .502}0.930 & -     & \cellcolor[rgb]{ .988,  .922,  .518}0.016\% & \cellcolor[rgb]{ .847,  .878,  .51}1.76\% & \cellcolor[rgb]{ 1,  .894,  .514}2.05\% & \cellcolor[rgb]{ .969,  .914,  .518}1.265 & 83 \\
    \midrule
    \textbf{RP} & \cellcolor[rgb]{ .996,  .882,  .51}0.094\% & \cellcolor[rgb]{ .91,  .894,  .51}1.319\% & \cellcolor[rgb]{ .996,  .886,  .51}7.10\% & \cellcolor[rgb]{ .788,  .863,  .506}-0.306 & \cellcolor[rgb]{ .467,  .765,  .486}5.02\% & \cellcolor[rgb]{ .996,  .863,  .506}0.931 & \cellcolor[rgb]{ .388,  .745,  .482}0.02 & \cellcolor[rgb]{ .996,  .863,  .506}0.010\% & \cellcolor[rgb]{ .992,  .843,  .502}-1.50\% & \cellcolor[rgb]{ .871,  .882,  .51}1.81\% & \cellcolor[rgb]{ .996,  .902,  .514}1.256 & 83 \\
    \midrule
    \textbf{MDP} & \cellcolor[rgb]{ .988,  .922,  .518}0.102\% & \cellcolor[rgb]{ .776,  .855,  .502}1.269\% & \cellcolor[rgb]{ .855,  .882,  .51}8.01\% & \cellcolor[rgb]{ .996,  .918,  .514}-0.316 & \cellcolor[rgb]{ .388,  .745,  .482}4.69\% & \cellcolor[rgb]{ .549,  .792,  .494}1.017 & \cellcolor[rgb]{ .62,  .812,  .494}0.15 & \cellcolor[rgb]{ .925,  .902,  .514}0.031\% & \cellcolor[rgb]{ .953,  .91,  .518}0.10\% & \cellcolor[rgb]{ .639,  .816,  .494}1.63\% & \cellcolor[rgb]{ .792,  .863,  .506}1.291 & 20 \\
    \midrule
    \textbf{Sust\_1} & \cellcolor[rgb]{ .973,  .412,  .42}0.044\% & \cellcolor[rgb]{ .396,  .745,  .482}1.127\% & \cellcolor[rgb]{ .973,  .412,  .42}3.90\% & \cellcolor[rgb]{ .62,  .812,  .498}-0.299 & \cellcolor[rgb]{ .929,  .898,  .51}6.94\% & \cellcolor[rgb]{ .973,  .412,  .42}0.896 & \cellcolor[rgb]{ .643,  .816,  .494}0.16 & \cellcolor[rgb]{ .973,  .412,  .42}-0.014\% & \cellcolor[rgb]{ .973,  .412,  .42}-5.96\% & \cellcolor[rgb]{ .451,  .761,  .482}1.48\% & \cellcolor[rgb]{ .973,  .412,  .42}1.139 & 18 \\
    \midrule
    \textbf{Sust\_2} & \cellcolor[rgb]{ .98,  .62,  .459}0.066\% & \cellcolor[rgb]{ .584,  .8,  .49}1.197\% & \cellcolor[rgb]{ .984,  .651,  .463}5.52\% & \cellcolor[rgb]{ .388,  .745,  .482}-0.289 & \cellcolor[rgb]{ .6,  .804,  .494}5.58\% & \cellcolor[rgb]{ .976,  .918,  .518}0.941 & \cellcolor[rgb]{ .922,  .898,  .51}0.32 & \cellcolor[rgb]{ .984,  .671,  .467}0.0001\% & \cellcolor[rgb]{ .98,  .584,  .451}-4.14\% & \cellcolor[rgb]{ .627,  .812,  .494}1.62\% & \cellcolor[rgb]{ .984,  .639,  .463}1.194 & 18 \\
    \midrule
    \textbf{Sust\_3} & \cellcolor[rgb]{ .765,  .855,  .506}0.170\% & \cellcolor[rgb]{ .992,  .71,  .478}1.983\% & \cellcolor[rgb]{ .741,  .847,  .506}8.58\% & \cellcolor[rgb]{ .988,  .722,  .478}-0.364 & \cellcolor[rgb]{ .992,  .745,  .486}9.51\% & \cellcolor[rgb]{ .753,  .851,  .506}0.981 & \cellcolor[rgb]{ .973,  .412,  .42}0.59 & \cellcolor[rgb]{ .788,  .863,  .506}0.064\% & \cellcolor[rgb]{ .588,  .804,  .494}5.95\% & \cellcolor[rgb]{ .988,  .671,  .471}3.19\% & \cellcolor[rgb]{ .82,  .871,  .51}1.287 & 11 \\
    \midrule
    \textbf{Sust\_4} & \cellcolor[rgb]{ .404,  .753,  .486}0.281\% & \cellcolor[rgb]{ .976,  .455,  .427}2.738\% & \cellcolor[rgb]{ .388,  .745,  .482}10.25\% & \cellcolor[rgb]{ .973,  .412,  .42}-0.440 & \cellcolor[rgb]{ .976,  .463,  .431}13.13\% & \cellcolor[rgb]{ .4,  .749,  .486}1.043 & \cellcolor[rgb]{ .988,  .667,  .471}0.48 & \cellcolor[rgb]{ .408,  .753,  .486}0.152\% & \cellcolor[rgb]{ .388,  .745,  .482}9.10\% & \cellcolor[rgb]{ .973,  .412,  .42}4.47\% & \cellcolor[rgb]{ .388,  .745,  .482}1.348 & 7 \\
    \midrule
    \textbf{Sust$\_{\boldsymbol{1}\_k}$}\textbf{Worst} & \cellcolor[rgb]{ .976,  .549,  .443}0.059\% & \cellcolor[rgb]{ .769,  .855,  .502}1.266\% & \cellcolor[rgb]{ .976,  .518,  .439}4.62\% & \cellcolor[rgb]{ .918,  .898,  .514}-0.312 & \cellcolor[rgb]{ 1,  .902,  .514}7.52\% & \cellcolor[rgb]{ .98,  .561,  .447}0.908 & \cellcolor[rgb]{ 1,  .922,  .518}0.36 & \cellcolor[rgb]{ .973,  .482,  .431}-0.010\% & \cellcolor[rgb]{ .976,  .514,  .439}-4.87\% & \cellcolor[rgb]{ .8,  .863,  .506}1.76\% & \cellcolor[rgb]{ .976,  .486,  .431}1.158 & 17 \\
    \midrule
    \textbf{Sust$\_{\boldsymbol{2}\_k}$}\textbf{Worst} & \cellcolor[rgb]{ .988,  .733,  .478}0.078\% & \cellcolor[rgb]{ 1,  .914,  .518}1.385\% & \cellcolor[rgb]{ .984,  .671,  .467}5.64\% & \cellcolor[rgb]{ .984,  .918,  .518}-0.314 & \cellcolor[rgb]{ 1,  .859,  .506}8.07\% & \cellcolor[rgb]{ .988,  .737,  .482}0.922 & \cellcolor[rgb]{ .992,  .714,  .478}0.46 & \cellcolor[rgb]{ .984,  .69,  .471}0.001\% & \cellcolor[rgb]{ .988,  .722,  .478}-2.74\% & \cellcolor[rgb]{ 1,  .902,  .518}2.01\% & \cellcolor[rgb]{ .98,  .612,  .455}1.188 & 15 \\
    \midrule
    \textbf{Sust$\_{\boldsymbol{3}\_k}$}\textbf{Worst} & \cellcolor[rgb]{ .725,  .843,  .502}0.182\% & \cellcolor[rgb]{ .988,  .655,  .467}2.144\% & \cellcolor[rgb]{ .761,  .855,  .506}8.48\% & \cellcolor[rgb]{ .984,  .675,  .467}-0.375 & \cellcolor[rgb]{ .984,  .612,  .459}11.22\% & \cellcolor[rgb]{ .745,  .851,  .506}0.982 & \cellcolor[rgb]{ .976,  .475,  .435}0.57 & \cellcolor[rgb]{ .753,  .851,  .506}0.071\% & \cellcolor[rgb]{ .584,  .804,  .494}5.97\% & \cellcolor[rgb]{ .988,  .643,  .467}3.32\% & \cellcolor[rgb]{ .89,  .89,  .514}1.277 & 9 \\
    \midrule
    \textbf{Sust$\_{\boldsymbol{4}\_k}$}\textbf{Worst} & \cellcolor[rgb]{ .388,  .745,  .482}0.285\% & \cellcolor[rgb]{ .973,  .412,  .42}2.858\% & \cellcolor[rgb]{ .447,  .765,  .486}9.98\% & \cellcolor[rgb]{ .973,  .431,  .424}-0.434 & \cellcolor[rgb]{ .973,  .412,  .42}13.79\% & \cellcolor[rgb]{ .388,  .745,  .482}1.045 & \cellcolor[rgb]{ .988,  .682,  .475}0.47 & \cellcolor[rgb]{ .388,  .745,  .482}0.156\% & \cellcolor[rgb]{ .42,  .757,  .486}8.62\% & \cellcolor[rgb]{ .976,  .443,  .427}4.33\% & \cellcolor[rgb]{ .49,  .776,  .49}1.334 & 6 \\
    \bottomrule
    \end{tabular}}%
  \label{tab:NASDAQ100_Perf_q1}%
\end{table}%

\noindent
%The same behavior is confirmed by observing the out-of-sample ROI over a three-years time horizon (see Table \ref{tab:NASDAQ100_ROI_q1}).
%
% Table generated by Excel2LaTeX from sheet 'ROI750'
\begin{table}[htbp!]
  \centering
  \caption{ROI based on a 3-year time horizon (with $k = 1$)}
\scalebox{0.7}{\begin{tabular}{|c|c|c|c|c|c|c|c|}
    \toprule
    \textbf{Approach} & \textbf{ExpRet} & \textbf{Vol} & \textbf{5\%-perc} & \textbf{25\%-perc} & \textbf{50\%-perc} & \textbf{75\%-perc} & \textbf{95\%-perc} \\
    \midrule
    \textbf{MV0} & \cellcolor[rgb]{ .973,  .412,  .42}34\% & \cellcolor[rgb]{ .388,  .745,  .482}9\% & \cellcolor[rgb]{ .973,  .412,  .42}22\% & \cellcolor[rgb]{ .973,  .412,  .42}26\% & \cellcolor[rgb]{ .973,  .412,  .42}33\% & \cellcolor[rgb]{ .973,  .412,  .42}40\% & \cellcolor[rgb]{ .973,  .412,  .42}52\% \\
    \midrule
    \textbf{EW} & \cellcolor[rgb]{ .992,  .922,  .518}121\% & \cellcolor[rgb]{ 1,  .918,  .518}18\% & \cellcolor[rgb]{ .988,  .918,  .518}100\% & \cellcolor[rgb]{ .992,  .922,  .518}109\% & \cellcolor[rgb]{ .992,  .922,  .518}116\% & \cellcolor[rgb]{ .992,  .922,  .518}132\% & \cellcolor[rgb]{ .992,  .922,  .518}159\% \\
    \midrule
    \textbf{RP} & \cellcolor[rgb]{ .996,  .863,  .506}100\% & \cellcolor[rgb]{ .804,  .863,  .506}14\% & \cellcolor[rgb]{ .996,  .871,  .506}82\% & \cellcolor[rgb]{ .996,  .863,  .506}92\% & \cellcolor[rgb]{ .996,  .855,  .502}97\% & \cellcolor[rgb]{ .996,  .867,  .506}108\% & \cellcolor[rgb]{ .996,  .871,  .506}128\% \\
    \midrule
    \textbf{MDP} & \cellcolor[rgb]{ .996,  .922,  .518}117\% & \cellcolor[rgb]{ .776,  .855,  .502}14\% & \cellcolor[rgb]{ .996,  .922,  .518}94\% & \cellcolor[rgb]{ .992,  .922,  .518}108\% & \cellcolor[rgb]{ .992,  .922,  .518}117\% & \cellcolor[rgb]{ .996,  .922,  .518}124\% & \cellcolor[rgb]{ 1,  .922,  .518}144\% \\
    \midrule
    \textbf{Sust\_1} & \cellcolor[rgb]{ .973,  .42,  .42}35\% & \cellcolor[rgb]{ .427,  .753,  .482}10\% & \cellcolor[rgb]{ .973,  .412,  .42}22\% & \cellcolor[rgb]{ .973,  .412,  .42}27\% & \cellcolor[rgb]{ .973,  .424,  .42}35\% & \cellcolor[rgb]{ .973,  .412,  .42}40\% & \cellcolor[rgb]{ .973,  .42,  .42}54\% \\
    \midrule
    \textbf{Sust\_2} & \cellcolor[rgb]{ .98,  .616,  .459}64\% & \cellcolor[rgb]{ .631,  .812,  .494}12\% & \cellcolor[rgb]{ .984,  .647,  .463}52\% & \cellcolor[rgb]{ .98,  .62,  .459}57\% & \cellcolor[rgb]{ .98,  .592,  .455}60\% & \cellcolor[rgb]{ .98,  .604,  .455}69\% & \cellcolor[rgb]{ .984,  .663,  .467}94\% \\
    \midrule
    \textbf{Sust\_3} & \cellcolor[rgb]{ .859,  .882,  .51}287\% & \cellcolor[rgb]{ .996,  .835,  .502}55\% & \cellcolor[rgb]{ .839,  .875,  .51}227\% & \cellcolor[rgb]{ .851,  .878,  .51}246\% & \cellcolor[rgb]{ .851,  .882,  .51}270\% & \cellcolor[rgb]{ .863,  .882,  .51}315\% & \cellcolor[rgb]{ .875,  .886,  .514}414\% \\
    \midrule
    \textbf{Sust\_4} & \cellcolor[rgb]{ .42,  .757,  .486}825\% & \cellcolor[rgb]{ .976,  .475,  .435}215\% & \cellcolor[rgb]{ .388,  .745,  .482}603\% & \cellcolor[rgb]{ .412,  .753,  .486}668\% & \cellcolor[rgb]{ .42,  .757,  .486}747\% & \cellcolor[rgb]{ .435,  .761,  .486}929\% & \cellcolor[rgb]{ .439,  .761,  .486}1344\% \\
    \midrule
    \textbf{Sust$\_{\boldsymbol{1}\_k}$}\textbf{Worst} & \cellcolor[rgb]{ .973,  .478,  .431}44\% & \cellcolor[rgb]{ .784,  .859,  .502}14\% & \cellcolor[rgb]{ .973,  .475,  .431}30\% & \cellcolor[rgb]{ .973,  .463,  .427}34\% & \cellcolor[rgb]{ .973,  .447,  .424}39\% & \cellcolor[rgb]{ .976,  .514,  .439}55\% & \cellcolor[rgb]{ .976,  .533,  .439}72\% \\
    \midrule
    \textbf{Sust$\_{\boldsymbol{2}\_k}$}\textbf{Worst} & \cellcolor[rgb]{ .984,  .667,  .467}72\% & \cellcolor[rgb]{ 1,  .918,  .518}19\% & \cellcolor[rgb]{ .98,  .627,  .459}50\% & \cellcolor[rgb]{ .98,  .608,  .455}55\% & \cellcolor[rgb]{ .984,  .655,  .467}69\% & \cellcolor[rgb]{ .984,  .698,  .475}83\% & \cellcolor[rgb]{ .992,  .78,  .49}113\% \\
    \midrule
    \textbf{Sust$\_{\boldsymbol{3}\_k}$}\textbf{Worst} & \cellcolor[rgb]{ .831,  .875,  .51}321\% & \cellcolor[rgb]{ .992,  .776,  .49}81\% & \cellcolor[rgb]{ .835,  .875,  .51}229\% & \cellcolor[rgb]{ .835,  .875,  .51}260\% & \cellcolor[rgb]{ .827,  .875,  .51}298\% & \cellcolor[rgb]{ .827,  .875,  .51}365\% & \cellcolor[rgb]{ .827,  .875,  .51}511\% \\
    \midrule
    \textbf{Sust$\_{\boldsymbol{4}\_k}$}\textbf{Worst} & \cellcolor[rgb]{ .388,  .745,  .482}862\% & \cellcolor[rgb]{ .973,  .412,  .42}243\% & \cellcolor[rgb]{ .396,  .749,  .486}599\% & \cellcolor[rgb]{ .388,  .745,  .482}690\% & \cellcolor[rgb]{ .388,  .745,  .482}777\% & \cellcolor[rgb]{ .388,  .745,  .482}992\% & \cellcolor[rgb]{ .388,  .745,  .482}1453\% \\
    \bottomrule
    \end{tabular}}%
  \label{tab:NASDAQ100_ROI_q1}%
\end{table}%
\noindent From the analysis of all the results, including those for $k=2,3,4$ reported in the Appendix, for the EuroStoxx50 data set, the best model seems to be the one adopting $k=3$, with a performance that improves when high target level for the portfolio expected return is fixed. This supports the idea that when there is  disagreement in the ESG ratings, including the evaluation of more than one agency provides an effective tool for improving the performance of the selected portfolio.

%%%%%%%%%%%%%%%%%%%%%%%%%%%%%%%%%%%%%%%%%%%%%%%%%%%%%%%%%%%%%%%%%%%%%%%%%%%%%%%%%%%%%%%%%%%%%%%%%%%%%%%%%%%%%%%%%%%%%%%%%%%%%%%%
\section{Conclusions}\label{sec:Conclusions}
In this paper we introduce a new optimization approach to portfolio selection. We start from the classical Markowitz framework and include the ESG evaluation criterion to obtain a three-objective portfolio optimization model which can be applied by practitioners from the various financial institutions. Security analysts and portfolio managers can apply this model in the classical and well-assessed mean-variance approach for their portfolio choice, but with the additional possibility of including sustainability issues in the selection process. The ESG objective of the model is formulated exploiting the $k-$sum operator which, for each stock in the market, allows to include $k\geq 1$ ESG evaluations provided by different agencies. This is a main issue when there is a significant disagreement among scores given by the different agencies to the same asset. We formulate the model as a convex quadratic program which minimizes the portfolio volatility. We provide the empirical construction of the three-objective efficient frontier, as well as, an extensive experimental analysis of the out-of-sample performance of optimal portfolios found by our model under suitable choices of target values for the ESG score and the expected return of the portfolio. It is worth nothing that our approach goes beyond the specific financial application context, and provides a general modelling framework which, in fact, extends the use of a $k-$sum optimization strategy from Linear to Quadratic Programming.
The study is motivated by the increasing importance of the sustainability criteria in portfolio selection problems and by the fact that security markets may be characterized by misalignment between ESG ratings provided by different agencies for the same asset. Our approach is able to make such different scores uniform and include all in a single measure of portfolio sustainability. Our model can be applied to any financial market and it easily adapts to any new scenarios given by multiple agencies evaluations; it is able to exploit all the available information on the ESG evaluation, but, if necessary, it can also select only a subset of ratings to be included in the analysis. Globally it is versatile, effective and computationally tractable.
We believe that all these features are crucial nowadays for an informed, transparent, and careful portfolio selection process, especially in view of the increasing importance of the ESG evaluation and corporate sustainability criteria also promoted by the European Commission action plan on sustainable finance to orient capitals towards sustainable investments.

%%%%%%%%%%%%%%%%%%%%%%%%%%%%%%%%%%%%%%%%%%%%%%%%%%%%%%%%%%%%%%%%%%%%%%%%%%%%%%%%%%%%%%%%%%%%%%%%%%%%%%%%%%%%%%%%%%%%%%%%%%%%%%%%
{\footnotesize
\bibliographystyle{apa}
\bibliography{Bibbase_MinMaxESG-PortfolioSelection_20230831}
}

%%%%%%%%%%%%%%%%%%%%%%%%%%%%%%%%%%%%%%%%%%%%%%%%%%%%%%%%%%%%%%%%%%%%%%%%%%%%%%%%%%%%%%%%%%%%%%%%%%%%%%%%%%%%%%%%%%%%%%%%%%%%%%%%
\section*{Appendix: Supplementary materials}\label{sec:Appendix}
For the sake of completeness, we provide additional tables containing the computational results obtained by the portfolio strategies listed in Table \ref{tab:ListOfModels} on the data sets considered in this study (see Table \ref{tab:DailyDatasets}).
More precisely, we report the cases with $k=2,3,4$ in our three-objective portfolio optimization model \eqref{eq:Triobjective_lambda_scalarized}.
We recall that if $k = 4$, we minimize the sum of all scores, while for $1 < k < 4$, we have intermediate worst-case levels.
For these purposes, in Tables \ref{tab:EuroStoxx50_Perf_q2},\ref{tab:EuroStoxx50_Perf_q3},\ref{tab:EuroStoxx50_Perf_q4},\ref{tab:EuroStoxx50_ROI_q2},\ref{tab:EuroStoxx50_ROI_q3},\ref{tab:EuroStoxx50_ROI_q4} we provide the computational results obtained with the EuroStoxx50 data set, while in tables \ref{tab:NASDAQ100_Perf_q2},\ref{tab:NASDAQ100_Perf_q3},\ref{tab:NASDAQ100_Perf_q4},\ref{tab:NASDAQ100_ROI_q2},\ref{tab:NASDAQ100_ROI_q3},\ref{tab:NASDAQ100_ROI_q4} we report the computational results obtained with the NASDAQ100 data set.

% Table generated by Excel2LaTeX from sheet 'Foglio1 (2)'
\begin{table}[htbp]
  \centering
  \caption{EuroStoxx50 (with $k = 2$)}
  \scalebox{0.7}{\hskip-1.65cm
\begin{tabular}{|c|c|c|c|c|c|c|c|c|c|c|c|c|}
    \toprule
    \textbf{Approach} & \textbf{ExpRet} & \textbf{Vol} & \textbf{Sharpe} & \textbf{MDD} & \textbf{Ulcer} & \textbf{Rachev10} & \textbf{Turn} & \textbf{AlphaJ} & \textbf{InfoRatio} & \textbf{VaR5} & \textbf{Omega} & \textbf{ave \#} \\
    \midrule
    \textbf{MV0} & \cellcolor[rgb]{ .984,  .694,  .471}0.042\% & \cellcolor[rgb]{ .388,  .745,  .482}0.939\% & \cellcolor[rgb]{ .992,  .812,  .494}4.45\% & \cellcolor[rgb]{ .996,  .875,  .506}-0.321 & \cellcolor[rgb]{ .996,  .816,  .498}8.53\% & \cellcolor[rgb]{ .827,  .875,  .51}0.918 & \cellcolor[rgb]{ .643,  .816,  .494}0.13 & \cellcolor[rgb]{ .988,  .733,  .478}0.025\% & \cellcolor[rgb]{ .98,  .627,  .459}2.46\% & \cellcolor[rgb]{ .388,  .745,  .482}1.22\% & \cellcolor[rgb]{ .992,  .839,  .498}1.145 & 12 \\
    \midrule
    \textbf{EW} & \cellcolor[rgb]{ .984,  .663,  .467}0.040\% & \cellcolor[rgb]{ .988,  .655,  .467}1.297\% & \cellcolor[rgb]{ .973,  .459,  .427}3.11\% & \cellcolor[rgb]{ .973,  .412,  .42}-0.407 & \cellcolor[rgb]{ .984,  .565,  .451}9.44\% & \cellcolor[rgb]{ .788,  .863,  .506}0.920 & -     & \cellcolor[rgb]{ .973,  .471,  .431}0.014\% & \cellcolor[rgb]{ .502,  .78,  .49}7.29\% & \cellcolor[rgb]{ .996,  .8,  .494}1.78\% & \cellcolor[rgb]{ .976,  .502,  .435}1.105 & 44 \\
    \midrule
    \textbf{RP} & \cellcolor[rgb]{ .984,  .639,  .463}0.039\% & \cellcolor[rgb]{ .996,  .788,  .494}1.201\% & \cellcolor[rgb]{ .976,  .498,  .435}3.26\% & \cellcolor[rgb]{ .976,  .518,  .439}-0.387 & \cellcolor[rgb]{ .992,  .745,  .486}8.79\% & \cellcolor[rgb]{ .992,  .8,  .494}0.905 & \cellcolor[rgb]{ .388,  .745,  .482}0.02 & \cellcolor[rgb]{ .976,  .49,  .431}0.015\% & \cellcolor[rgb]{ .388,  .745,  .482}7.77\% & \cellcolor[rgb]{ 1,  .918,  .518}1.60\% & \cellcolor[rgb]{ .976,  .545,  .443}1.110 & 44 \\
\bottomrule
    \end{tabular}}%
  \label{tab:EuroStoxx50_Perf_q2}%
\end{table}%
%%%%%%%%%%%%%%%%%%%%%%%%%%%%%%%%%%%%%%%%%%%%%%%%%%%%%%%%%%%%%%%%%%%%%%%%%%%%%%%%%%%%%%%%%%%%%%%%%%%%%%%%%%%%%%%%%%%%%%%%%%%%
% Table generated by Excel2LaTeX from sheet 'Foglio1 (2)'
\begin{table}[htbp]
  \centering
  \caption{EuroStoxx50 (with $k = 3$)}\scalebox{0.7}{\hskip-1.65cm
    \begin{tabular}{|c|c|c|c|c|c|c|c|c|c|c|c|c|}
    \toprule
    \textbf{Approach} & \textbf{ExpRet} & \textbf{Vol} & \textbf{Sharpe} & \textbf{MDD} & \textbf{Ulcer} & \textbf{Rachev10} & \textbf{Turn} & \textbf{AlphaJ} & \textbf{InfoRatio} & \textbf{VaR5} & \textbf{Omega} & \textbf{ave \#} \\
    \midrule
    \textbf{MV0} & \cellcolor[rgb]{ .984,  .694,  .471}0.042\% & \cellcolor[rgb]{ .388,  .745,  .482}0.939\% & \cellcolor[rgb]{ .992,  .812,  .494}4.45\% & \cellcolor[rgb]{ .996,  .875,  .506}-0.321 & \cellcolor[rgb]{ 1,  .871,  .51}8.53\% & \cellcolor[rgb]{ .996,  .91,  .514}0.918 & \cellcolor[rgb]{ .686,  .831,  .498}0.13 & \cellcolor[rgb]{ .988,  .733,  .478}0.025\% & \cellcolor[rgb]{ .984,  .639,  .463}2.46\% & \cellcolor[rgb]{ .388,  .745,  .482}1.22\% & \cellcolor[rgb]{ .992,  .839,  .498}1.145 & 12 \\
    \midrule
    \textbf{EW} & \cellcolor[rgb]{ .984,  .663,  .467}0.040\% & \cellcolor[rgb]{ .984,  .584,  .455}1.297\% & \cellcolor[rgb]{ .973,  .459,  .427}3.11\% & \cellcolor[rgb]{ .973,  .412,  .42}-0.407 & \cellcolor[rgb]{ .984,  .588,  .455}9.44\% & \cellcolor[rgb]{ .965,  .914,  .518}0.920 & -     & \cellcolor[rgb]{ .973,  .471,  .431}0.014\% & \cellcolor[rgb]{ .49,  .776,  .49}7.29\% & \cellcolor[rgb]{ .992,  .745,  .486}1.78\% & \cellcolor[rgb]{ .976,  .502,  .435}1.105 & 44 \\
    \midrule
    \textbf{RP} & \cellcolor[rgb]{ .984,  .639,  .463}0.039\% & \cellcolor[rgb]{ .992,  .729,  .482}1.201\% & \cellcolor[rgb]{ .976,  .498,  .435}3.26\% & \cellcolor[rgb]{ .976,  .518,  .439}-0.387 & \cellcolor[rgb]{ .996,  .788,  .494}8.79\% & \cellcolor[rgb]{ .988,  .718,  .478}0.905 & \cellcolor[rgb]{ .388,  .745,  .482}0.02 & \cellcolor[rgb]{ .976,  .49,  .431}0.015\% & \cellcolor[rgb]{ .388,  .745,  .482}7.77\% & \cellcolor[rgb]{ 1,  .878,  .51}1.60\% & \cellcolor[rgb]{ .976,  .545,  .443}1.110 & 44 \\
\bottomrule
    \end{tabular}}%
  \label{tab:EuroStoxx50_Perf_q3}%
\end{table}%
%%%%%%%%%%%%%%%%%%%%%%%%%%%%%%%%%%%%%%%%%%%%%%%%%%%%%%%%%%%%%%%%%%%%%%%%%%%%%%%%%%%%%%%%%%%%%%%%%%%%%%%%%%%%%%%%%%%%%%%%%%%%
% Table generated by Excel2LaTeX from sheet 'Foglio1 (2)'
\begin{table}[htbp]
  \centering
  \caption{EuroStoxx50 (with $k = 4$)}\scalebox{0.7}{\hskip-1.65cm
    \begin{tabular}{|c|c|c|c|c|c|c|c|c|c|c|c|c|}
    \toprule
    \textbf{Approach} & \textbf{ExpRet} & \textbf{Vol} & \textbf{Sharpe} & \textbf{MDD} & \textbf{Ulcer} & \textbf{Rachev10} & \textbf{Turn} & \textbf{AlphaJ} & \textbf{InfoRatio} & \textbf{VaR5} & \textbf{Omega} & \textbf{ave \#} \\
    \midrule
    \textbf{MV0} & \cellcolor[rgb]{ .984,  .694,  .471}0.042\% & \cellcolor[rgb]{ .388,  .745,  .482}0.939\% & \cellcolor[rgb]{ .992,  .812,  .494}4.45\% & \cellcolor[rgb]{ .996,  .875,  .506}-0.321 & \cellcolor[rgb]{ 1,  .875,  .51}8.53\% & \cellcolor[rgb]{ .996,  .898,  .51}0.918 & \cellcolor[rgb]{ .651,  .82,  .494}0.13 & \cellcolor[rgb]{ .988,  .733,  .478}0.025\% & \cellcolor[rgb]{ .984,  .631,  .459}2.46\% & \cellcolor[rgb]{ .388,  .745,  .482}1.22\% & \cellcolor[rgb]{ .992,  .839,  .498}1.145 & 12 \\
    \midrule
    \textbf{EW} & \cellcolor[rgb]{ .984,  .663,  .467}0.040\% & \cellcolor[rgb]{ .992,  .714,  .478}1.297\% & \cellcolor[rgb]{ .973,  .459,  .427}3.11\% & \cellcolor[rgb]{ .973,  .412,  .42}-0.407 & \cellcolor[rgb]{ .984,  .588,  .455}9.44\% & \cellcolor[rgb]{ .996,  .918,  .514}0.920 & -     & \cellcolor[rgb]{ .973,  .471,  .431}0.014\% & \cellcolor[rgb]{ .494,  .776,  .49}7.29\% & \cellcolor[rgb]{ .996,  .788,  .494}1.78\% & \cellcolor[rgb]{ .976,  .502,  .435}1.105 & 44 \\
    \midrule
    \textbf{RP} & \cellcolor[rgb]{ .984,  .639,  .463}0.039\% & \cellcolor[rgb]{ .996,  .847,  .506}1.201\% & \cellcolor[rgb]{ .976,  .498,  .435}3.26\% & \cellcolor[rgb]{ .976,  .518,  .439}-0.387 & \cellcolor[rgb]{ .996,  .792,  .494}8.79\% & \cellcolor[rgb]{ .988,  .714,  .475}0.905 & \cellcolor[rgb]{ .388,  .745,  .482}0.02 & \cellcolor[rgb]{ .976,  .49,  .431}0.015\% & \cellcolor[rgb]{ .388,  .745,  .482}7.77\% & \cellcolor[rgb]{ 1,  .894,  .514}1.60\% & \cellcolor[rgb]{ .976,  .545,  .443}1.110 & 44 \\
\bottomrule
    \end{tabular}}%
  \label{tab:EuroStoxx50_Perf_q4}%
\end{table}%
%%%%%%%%%%%%%%%%%%%%%%%%%%%%%%%%%%%%%%%%%%%%%%%%%%%%%%%%%%%%%%%%%%%%%%%%%%%%%%%%%%%%%%%%%%%%%%%%%%%%%%%%%%%%%%%%%%%%%%%%%%%%
% Table generated by Excel2LaTeX from sheet 'ROI750 (2)'
\begin{table}[htbp]
  \centering
  \caption{ROI based on a 3-year time horizon (with $k = 2$)}
    \scalebox{0.7}{\begin{tabular}{|l|c|c|c|c|c|c|c|c|c|}
    \toprule
    \multicolumn{1}{|c|}{\textbf{Approach}} & \textbf{ExpRet} & \textbf{Vol} & \textbf{5\%-perc} & \textbf{25\%-perc} & \textbf{50\%-perc} & \textbf{75\%-perc} & \textbf{95\%-perc} & \textbf{Sharpe} & \textbf{InfoRatio} \\
    \midrule
    \multicolumn{1}{|c|}{\textbf{MV0}} & \cellcolor[rgb]{ .98,  .588,  .451}21\% & \cellcolor[rgb]{ .812,  .867,  .506}11\% & \cellcolor[rgb]{ .98,  .612,  .455}9\% & \cellcolor[rgb]{ .976,  .533,  .443}13\% & \cellcolor[rgb]{ .976,  .541,  .443}17\% & \cellcolor[rgb]{ .98,  .627,  .459}30\% & \cellcolor[rgb]{ .984,  .659,  .467}41\% & \cellcolor[rgb]{ .976,  .51,  .435}202\% & \cellcolor[rgb]{ .98,  .573,  .451}73\% \\
    \midrule
    \multicolumn{1}{|c|}{\textbf{EW}} & \cellcolor[rgb]{ .988,  .761,  .486}30\% & \cellcolor[rgb]{ .996,  .804,  .498}16\% & \cellcolor[rgb]{ .973,  .412,  .42}1\% & \cellcolor[rgb]{ .988,  .714,  .475}19\% & \cellcolor[rgb]{ .996,  .882,  .51}31\% & \cellcolor[rgb]{ .992,  .827,  .498}43\% & \cellcolor[rgb]{ .996,  .859,  .502}56\% & \cellcolor[rgb]{ .973,  .412,  .42}190\% & \cellcolor[rgb]{ .529,  .788,  .494}749\% \\
    \midrule
    \multicolumn{1}{|c|}{\textbf{RP}} & \cellcolor[rgb]{ .988,  .725,  .478}28\% & \cellcolor[rgb]{ 1,  .878,  .51}14\% & \cellcolor[rgb]{ .976,  .486,  .431}4\% & \cellcolor[rgb]{ .984,  .694,  .475}19\% & \cellcolor[rgb]{ .992,  .804,  .494}28\% & \cellcolor[rgb]{ .992,  .776,  .49}39\% & \cellcolor[rgb]{ .992,  .796,  .49}51\% & \cellcolor[rgb]{ .976,  .51,  .435}202\% & \cellcolor[rgb]{ .388,  .745,  .482}879\% \\
\bottomrule
    \end{tabular}}%
  \label{tab:EuroStoxx50_ROI_q2}%
\end{table}%
%%%%%%%%%%%%%%%%%%%%%%%%%%%%%%%%%%%%%%%%%%%%%%%%%%%%%%%%%%%%%%%%%%%%%%%%%%%%%%%%%%%%%%%%%%%%%%%%%%%%%%%%%%%%%%%%%%%%%%%%%%%%
% Table generated by Excel2LaTeX from sheet 'ROI750 (2)'
\begin{table}[htbp]
  \centering
  \caption{ROI based on a 3-year time horizon (with $k = 3$)}
    \scalebox{0.7}{\begin{tabular}{|l|c|c|c|c|c|c|c|c|c|}
    \toprule
    \multicolumn{1}{|c|}{\textbf{Approach}} & \textbf{ExpRet} & \textbf{Vol} & \textbf{5\%-perc} & \textbf{25\%-perc} & \textbf{50\%-perc} & \textbf{75\%-perc} & \textbf{95\%-perc} & \textbf{Sharpe} & \textbf{InfoRatio} \\
    \midrule
    \multicolumn{1}{|c|}{\textbf{MV0}} & \cellcolor[rgb]{ .98,  .588,  .451}21\% & \cellcolor[rgb]{ .812,  .867,  .506}11\% & \cellcolor[rgb]{ .98,  .612,  .455}9\% & \cellcolor[rgb]{ .976,  .533,  .443}13\% & \cellcolor[rgb]{ .976,  .541,  .443}17\% & \cellcolor[rgb]{ .98,  .627,  .459}30\% & \cellcolor[rgb]{ .984,  .663,  .467}41\% & \cellcolor[rgb]{ .976,  .51,  .435}202\% & \cellcolor[rgb]{ .98,  .565,  .447}73\% \\
    \midrule
    \multicolumn{1}{|c|}{\textbf{EW}} & \cellcolor[rgb]{ .988,  .761,  .486}30\% & \cellcolor[rgb]{ .996,  .804,  .498}16\% & \cellcolor[rgb]{ .973,  .412,  .42}1\% & \cellcolor[rgb]{ .988,  .714,  .475}19\% & \cellcolor[rgb]{ .996,  .882,  .51}31\% & \cellcolor[rgb]{ .992,  .827,  .498}43\% & \cellcolor[rgb]{ .996,  .867,  .506}56\% & \cellcolor[rgb]{ .973,  .412,  .42}190\% & \cellcolor[rgb]{ .533,  .788,  .494}749\% \\
    \midrule
    \multicolumn{1}{|c|}{\textbf{RP}} & \cellcolor[rgb]{ .988,  .725,  .478}28\% & \cellcolor[rgb]{ 1,  .875,  .51}14\% & \cellcolor[rgb]{ .976,  .486,  .431}4\% & \cellcolor[rgb]{ .984,  .694,  .475}19\% & \cellcolor[rgb]{ .992,  .804,  .494}28\% & \cellcolor[rgb]{ .992,  .776,  .486}39\% & \cellcolor[rgb]{ .992,  .804,  .494}51\% & \cellcolor[rgb]{ .976,  .51,  .435}202\% & \cellcolor[rgb]{ .388,  .745,  .482}879\% \\
\bottomrule
    \end{tabular}}%
  \label{tab:EuroStoxx50_ROI_q3}%
\end{table}%
%%%%%%%%%%%%%%%%%%%%%%%%%%%%%%%%%%%%%%%%%%%%%%%%%%%%%%%%%%%%%%%%%%%%%%%%%%%%%%%%%%%%%%%%%%%%%%%%%%%%%%%%%%%%%%%%%%%%%%%%%%%%
% Table generated by Excel2LaTeX from sheet 'ROI750 (2)'
\begin{table}[htbp]
  \centering
  \caption{ROI based on a 3-year time horizon (with $k = 4$)}
    \scalebox{0.7}{\begin{tabular}{|c|c|c|c|c|c|c|c|c|c|}
    \toprule
    \textbf{Approach} & \textbf{ExpRet} & \textbf{Vol} & \textbf{5\%-perc} & \textbf{25\%-perc} & \textbf{50\%-perc} & \textbf{75\%-perc} & \textbf{95\%-perc} & \textbf{Sharpe} & \textbf{InfoRatio} \\
    \midrule
    \textbf{MV0} & \cellcolor[rgb]{ .98,  .588,  .451}21\% & \cellcolor[rgb]{ .812,  .867,  .506}11\% & \cellcolor[rgb]{ .98,  .612,  .455}9\% & \cellcolor[rgb]{ .976,  .533,  .443}13\% & \cellcolor[rgb]{ .976,  .541,  .443}17\% & \cellcolor[rgb]{ .98,  .624,  .459}30\% & \cellcolor[rgb]{ .984,  .647,  .463}41\% & \cellcolor[rgb]{ .976,  .518,  .439}202\% & \cellcolor[rgb]{ .98,  .565,  .447}73\% \\
    \midrule
    \textbf{EW} & \cellcolor[rgb]{ .988,  .761,  .486}30\% & \cellcolor[rgb]{ .996,  .792,  .494}16\% & \cellcolor[rgb]{ .973,  .412,  .42}1\% & \cellcolor[rgb]{ .988,  .714,  .475}19\% & \cellcolor[rgb]{ .996,  .882,  .51}31\% & \cellcolor[rgb]{ .992,  .824,  .498}43\% & \cellcolor[rgb]{ .992,  .839,  .502}56\% & \cellcolor[rgb]{ .973,  .412,  .42}190\% & \cellcolor[rgb]{ .533,  .788,  .494}749\% \\
    \midrule
    \textbf{RP} & \cellcolor[rgb]{ .988,  .725,  .478}28\% & \cellcolor[rgb]{ 1,  .871,  .51}14\% & \cellcolor[rgb]{ .976,  .486,  .431}4\% & \cellcolor[rgb]{ .984,  .694,  .475}19\% & \cellcolor[rgb]{ .992,  .804,  .494}28\% & \cellcolor[rgb]{ .988,  .773,  .486}39\% & \cellcolor[rgb]{ .992,  .78,  .49}51\% & \cellcolor[rgb]{ .976,  .514,  .439}202\% & \cellcolor[rgb]{ .388,  .745,  .482}879\% \\
    \midrule
    \textbf{MDP} & \cellcolor[rgb]{ .996,  .918,  .514}38\% & \cellcolor[rgb]{ .902,  .89,  .51}12\% & \cellcolor[rgb]{ .949,  .91,  .518}25\% & \cellcolor[rgb]{ .969,  .914,  .518}28\% & \cellcolor[rgb]{ .996,  .914,  .514}33\% & \cellcolor[rgb]{ .996,  .902,  .514}48\% & \cellcolor[rgb]{ .996,  .878,  .51}59\% & \cellcolor[rgb]{ .878,  .886,  .514}324\% & \cellcolor[rgb]{ .996,  .914,  .514}323\% \\
\bottomrule
    \end{tabular}}%
  \label{tab:EuroStoxx50_ROI_q4}%
\end{table}%
%%%%%%%%%%%%%%%%%%%%%%%%%%%%%%%%%%%%%%%%%%%%%%%%%%%%%%%%%%%%%%%%%%%%%%%%%%%%%%%%%%%%%%%%%%%%%%%%%%%%%%%%%%%%%%%%%%%%%%%%%%%%
% Table generated by Excel2LaTeX from sheet 'Foglio1 (2)'
\begin{table}[htbp]
  \centering
  \caption{NASDAQ100 (with $k = 2$)}
    \scalebox{0.7}{\hskip-1.65cm \begin{tabular}{|c|c|c|c|c|c|c|c|c|c|c|c|c|}
    \toprule
    \textbf{Approach} & \textbf{ExpRet} & \textbf{Vol} & \textbf{Sharpe} & \textbf{MDD} & \textbf{Ulcer} & \textbf{Rachev10} & \textbf{Turn} & \textbf{AlphaJ} & \textbf{InfoRatio} & \textbf{VaR5} & \textbf{Omega} & \textbf{ave \#} \\
    \midrule
    \textbf{MV0} & \cellcolor[rgb]{ .973,  .42,  .42}0.045\% & \cellcolor[rgb]{ .388,  .745,  .482}1.124\% & \cellcolor[rgb]{ .973,  .424,  .42}3.99\% & \cellcolor[rgb]{ .996,  .902,  .514}-0.319 & \cellcolor[rgb]{ 1,  .89,  .514}6.83\% & \cellcolor[rgb]{ .973,  .455,  .427}0.901 & \cellcolor[rgb]{ .627,  .812,  .494}0.15 & \cellcolor[rgb]{ .973,  .439,  .424}-0.013\% & \cellcolor[rgb]{ .973,  .424,  .42}-5.80\% & \cellcolor[rgb]{ .388,  .745,  .482}1.43\% & \cellcolor[rgb]{ .973,  .435,  .424}1.145 & 19 \\
    \midrule
    \textbf{EW} & \cellcolor[rgb]{ .969,  .914,  .518}0.108\% & \cellcolor[rgb]{ 1,  .875,  .51}1.435\% & \cellcolor[rgb]{ .961,  .91,  .518}7.53\% & \cellcolor[rgb]{ .949,  .91,  .518}-0.312 & \cellcolor[rgb]{ .722,  .839,  .498}5.61\% & \cellcolor[rgb]{ .988,  .71,  .475}0.930 & -     & \cellcolor[rgb]{ .988,  .922,  .518}0.016\% & \cellcolor[rgb]{ .851,  .878,  .51}1.76\% & \cellcolor[rgb]{ 1,  .871,  .51}2.05\% & \cellcolor[rgb]{ .973,  .914,  .518}1.265 & 83 \\
    \midrule
    \textbf{RP} & \cellcolor[rgb]{ .996,  .882,  .51}0.094\% & \cellcolor[rgb]{ 1,  .914,  .518}1.319\% & \cellcolor[rgb]{ .996,  .886,  .51}7.10\% & \cellcolor[rgb]{ .812,  .867,  .51}-0.306 & \cellcolor[rgb]{ .506,  .776,  .486}5.02\% & \cellcolor[rgb]{ .988,  .725,  .478}0.931 & \cellcolor[rgb]{ .388,  .745,  .482}0.02 & \cellcolor[rgb]{ .996,  .863,  .506}0.010\% & \cellcolor[rgb]{ .992,  .843,  .502}-1.50\% & \cellcolor[rgb]{ 1,  .918,  .518}1.81\% & \cellcolor[rgb]{ .996,  .902,  .514}1.256 & 83 \\
\bottomrule
    \end{tabular}}%
  \label{tab:NASDAQ100_Perf_q2}%
\end{table}%
%%%%%%%%%%%%%%%%%%%%%%%%%%%%%%%%%%%%%%%%%%%%%%%%%%%%%%%%%%%%%%%%%%%%%%%%%%%%%%%%%%%%%%%%%%%%%%%%%%%%%%%%%%%%%%%%%%%%%%%%%%%%
% Table generated by Excel2LaTeX from sheet 'Foglio1 (2)'
\begin{table}[htbp]
  \centering
  \caption{NASDAQ100 (with $k = 3$)}
   \scalebox{0.7}{\hskip-1.65cm \begin{tabular}{|c|c|c|c|c|c|c|c|c|c|c|c|c|}
    \toprule
    \textbf{Approach} & \textbf{ExpRet} & \textbf{Vol} & \textbf{Sharpe} & \textbf{MDD} & \textbf{Ulcer} & \textbf{Rachev10} & \textbf{Turn} & \textbf{AlphaJ} & \textbf{InfoRatio} & \textbf{VaR5} & \textbf{Omega} & \textbf{ave \#} \\
    \midrule
    \textbf{MV0} & \cellcolor[rgb]{ .973,  .42,  .42}0.045\% & \cellcolor[rgb]{ .388,  .745,  .482}1.124\% & \cellcolor[rgb]{ .973,  .424,  .42}3.99\% & \cellcolor[rgb]{ .996,  .898,  .514}-0.319 & \cellcolor[rgb]{ 1,  .89,  .514}6.83\% & \cellcolor[rgb]{ .973,  .463,  .427}0.901 & \cellcolor[rgb]{ .655,  .82,  .494}0.15 & \cellcolor[rgb]{ .973,  .439,  .424}-0.013\% & \cellcolor[rgb]{ .973,  .424,  .42}-5.80\% & \cellcolor[rgb]{ .388,  .745,  .482}1.43\% & \cellcolor[rgb]{ .973,  .435,  .424}1.145 & 19 \\
    \midrule
    \textbf{EW} & \cellcolor[rgb]{ .969,  .914,  .518}0.108\% & \cellcolor[rgb]{ 1,  .875,  .51}1.435\% & \cellcolor[rgb]{ .961,  .91,  .518}7.53\% & \cellcolor[rgb]{ .949,  .91,  .518}-0.312 & \cellcolor[rgb]{ .725,  .839,  .498}5.61\% & \cellcolor[rgb]{ .988,  .745,  .482}0.930 & -     & \cellcolor[rgb]{ .988,  .922,  .518}0.016\% & \cellcolor[rgb]{ .851,  .878,  .51}1.76\% & \cellcolor[rgb]{ 1,  .871,  .51}2.05\% & \cellcolor[rgb]{ .973,  .914,  .518}1.265 & 83 \\
    \midrule
    \textbf{RP} & \cellcolor[rgb]{ .996,  .882,  .51}0.094\% & \cellcolor[rgb]{ 1,  .914,  .518}1.319\% & \cellcolor[rgb]{ .996,  .886,  .51}7.10\% & \cellcolor[rgb]{ .812,  .867,  .51}-0.306 & \cellcolor[rgb]{ .51,  .776,  .486}5.02\% & \cellcolor[rgb]{ .988,  .765,  .486}0.931 & \cellcolor[rgb]{ .388,  .745,  .482}0.02 & \cellcolor[rgb]{ .996,  .863,  .506}0.010\% & \cellcolor[rgb]{ .992,  .843,  .502}-1.50\% & \cellcolor[rgb]{ 1,  .914,  .518}1.81\% & \cellcolor[rgb]{ .996,  .902,  .514}1.256 & 83 \\
\bottomrule
    \end{tabular}}%
  \label{tab:NASDAQ100_Perf_q3}%
\end{table}%
%%%%%%%%%%%%%%%%%%%%%%%%%%%%%%%%%%%%%%%%%%%%%%%%%%%%%%%%%%%%%%%%%%%%%%%%%%%%%%%%%%%%%%%%%%%%%%%%%%%%%%%%%%%%%%%%%%%%%%%%%%%%
% Table generated by Excel2LaTeX from sheet 'Foglio1 (2)'
\begin{table}[htbp]
  \centering
  \caption{NASDAQ100 (with $k = 4$)}
   \scalebox{0.7}{\hskip-1.65cm \begin{tabular}{|c|c|c|c|c|c|c|c|c|c|c|c|c|}
    \toprule
    \textbf{Approach} & \textbf{ExpRet} & \textbf{Vol} & \textbf{Sharpe} & \textbf{MDD} & \textbf{Ulcer} & \textbf{Rachev10} & \textbf{Turn} & \textbf{AlphaJ} & \textbf{InfoRatio} & \textbf{VaR5} & \textbf{Omega} & \textbf{ave \#} \\
    \midrule
    \textbf{MV0} & \cellcolor[rgb]{ .973,  .42,  .42}0.045\% & \cellcolor[rgb]{ .388,  .745,  .482}1.124\% & \cellcolor[rgb]{ .973,  .424,  .42}3.99\% & \cellcolor[rgb]{ .996,  .898,  .514}-0.319 & \cellcolor[rgb]{ 1,  .91,  .518}6.83\% & \cellcolor[rgb]{ .973,  .463,  .427}0.901 & \cellcolor[rgb]{ .655,  .82,  .494}0.15 & \cellcolor[rgb]{ .973,  .439,  .424}-0.013\% & \cellcolor[rgb]{ .973,  .424,  .42}-5.80\% & \cellcolor[rgb]{ .388,  .745,  .482}1.43\% & \cellcolor[rgb]{ .973,  .435,  .424}1.145 & 19 \\
    \midrule
    \textbf{EW} & \cellcolor[rgb]{ .969,  .914,  .518}0.108\% & \cellcolor[rgb]{ 1,  .875,  .51}1.435\% & \cellcolor[rgb]{ .961,  .91,  .518}7.53\% & \cellcolor[rgb]{ .949,  .91,  .518}-0.312 & \cellcolor[rgb]{ .671,  .824,  .498}5.61\% & \cellcolor[rgb]{ .988,  .741,  .482}0.930 & -     & \cellcolor[rgb]{ .988,  .922,  .518}0.016\% & \cellcolor[rgb]{ .851,  .878,  .51}1.76\% & \cellcolor[rgb]{ 1,  .875,  .51}2.05\% & \cellcolor[rgb]{ .973,  .914,  .518}1.265 & 83 \\
    \midrule
    \textbf{RP} & \cellcolor[rgb]{ .996,  .882,  .51}0.094\% & \cellcolor[rgb]{ 1,  .914,  .518}1.319\% & \cellcolor[rgb]{ .996,  .886,  .51}7.10\% & \cellcolor[rgb]{ .812,  .867,  .51}-0.306 & \cellcolor[rgb]{ .49,  .773,  .486}5.02\% & \cellcolor[rgb]{ .988,  .761,  .486}0.931 & \cellcolor[rgb]{ .388,  .745,  .482}0.02 & \cellcolor[rgb]{ .996,  .863,  .506}0.010\% & \cellcolor[rgb]{ .992,  .843,  .502}-1.50\% & \cellcolor[rgb]{ 1,  .918,  .518}1.81\% & \cellcolor[rgb]{ .996,  .902,  .514}1.256 & 83 \\
\bottomrule
    \end{tabular}}%
  \label{tab:NASDAQ100_Perf_q4}%
\end{table}%
%%%%%%%%%%%%%%%%%%%%%%%%%%%%%%%%%%%%%%%%%%%%%%%%%%%%%%%%%%%%%%%%%%%%%%%%%%%%%%%%%%%%%%%%%%%%%%%%%%%%%%%%%%%%%%%%%%%%%%%%%%%%
% Table generated by Excel2LaTeX from sheet 'ROI750 (2)'
\begin{table}[htbp]
  \centering
  \caption{ROI based on a 3-year time horizon (with $k = 2$)}
    \scalebox{0.7}{\begin{tabular}{|c|c|c|c|c|c|c|c|}
    \toprule
    \textbf{Approach} & \textbf{ExpRet} & \textbf{Vol} & \textbf{5\%-perc} & \textbf{25\%-perc} & \textbf{50\%-perc} & \textbf{75\%-perc} & \textbf{95\%-perc} \\
    \midrule
    \textbf{MV0} & \cellcolor[rgb]{ .973,  .412,  .42}34\% & \cellcolor[rgb]{ .42,  .753,  .482}9\% & \cellcolor[rgb]{ .973,  .412,  .42}22\% & \cellcolor[rgb]{ .973,  .412,  .42}26\% & \cellcolor[rgb]{ .973,  .412,  .42}33\% & \cellcolor[rgb]{ .973,  .412,  .42}40\% & \cellcolor[rgb]{ .973,  .412,  .42}52\% \\
    \midrule
    \textbf{EW} & \cellcolor[rgb]{ .992,  .922,  .518}121\% & \cellcolor[rgb]{ 1,  .914,  .518}18\% & \cellcolor[rgb]{ .988,  .922,  .518}100\% & \cellcolor[rgb]{ .992,  .922,  .518}109\% & \cellcolor[rgb]{ .992,  .922,  .518}116\% & \cellcolor[rgb]{ .992,  .922,  .518}132\% & \cellcolor[rgb]{ .992,  .922,  .518}159\% \\
    \midrule
    \textbf{RP} & \cellcolor[rgb]{ .996,  .863,  .506}100\% & \cellcolor[rgb]{ .973,  .914,  .514}14\% & \cellcolor[rgb]{ .996,  .871,  .506}82\% & \cellcolor[rgb]{ .996,  .863,  .506}92\% & \cellcolor[rgb]{ .996,  .855,  .502}97\% & \cellcolor[rgb]{ .996,  .867,  .506}108\% & \cellcolor[rgb]{ .996,  .871,  .506}128\% \\
    \midrule
    \textbf{MDP} & \cellcolor[rgb]{ .996,  .922,  .518}117\% & \cellcolor[rgb]{ .941,  .902,  .514}14\% & \cellcolor[rgb]{ .996,  .922,  .518}94\% & \cellcolor[rgb]{ .992,  .922,  .518}108\% & \cellcolor[rgb]{ .992,  .922,  .518}117\% & \cellcolor[rgb]{ .996,  .922,  .518}124\% & \cellcolor[rgb]{ 1,  .922,  .518}144\% \\
    \midrule
    \textbf{Sust\_1} & \cellcolor[rgb]{ .973,  .42,  .42}35\% & \cellcolor[rgb]{ .471,  .769,  .486}10\% & \cellcolor[rgb]{ .973,  .412,  .42}22\% & \cellcolor[rgb]{ .973,  .412,  .42}27\% & \cellcolor[rgb]{ .973,  .424,  .42}35\% & \cellcolor[rgb]{ .973,  .412,  .42}40\% & \cellcolor[rgb]{ .973,  .42,  .42}54\% \\
\bottomrule
    \end{tabular}}%
  \label{tab:NASDAQ100_ROI_q2}%
\end{table}%
%%%%%%%%%%%%%%%%%%%%%%%%%%%%%%%%%%%%%%%%%%%%%%%%%%%%%%%%%%%%%%%%%%%%%%%%%%%%%%%%%%%%%%%%%%%%%%%%%%%%%%%%%%%%%%%%%%%%%%%%%%%%
% Table generated by Excel2LaTeX from sheet 'ROI750 (2)'
\begin{table}[htbp]
  \centering
  \caption{ROI based on a 3-year time horizon (with $k = 3$)}
    \scalebox{0.7}{\begin{tabular}{|c|c|c|c|c|c|c|c|}
    \toprule
    \textbf{Approach} & \textbf{ExpRet} & \textbf{Vol} & \textbf{5\%-perc} & \textbf{25\%-perc} & \textbf{50\%-perc} & \textbf{75\%-perc} & \textbf{95\%-perc} \\
    \midrule
    \textbf{MV0} & \cellcolor[rgb]{ .973,  .412,  .42}34\% & \cellcolor[rgb]{ .522,  .78,  .486}9\% & \cellcolor[rgb]{ .973,  .412,  .42}22\% & \cellcolor[rgb]{ .973,  .412,  .42}26\% & \cellcolor[rgb]{ .973,  .412,  .42}33\% & \cellcolor[rgb]{ .973,  .412,  .42}40\% & \cellcolor[rgb]{ .973,  .412,  .42}52\% \\
    \midrule
    \textbf{EW} & \cellcolor[rgb]{ .992,  .922,  .518}121\% & \cellcolor[rgb]{ 1,  .914,  .518}18\% & \cellcolor[rgb]{ .988,  .922,  .518}100\% & \cellcolor[rgb]{ .992,  .922,  .518}109\% & \cellcolor[rgb]{ .996,  .922,  .518}116\% & \cellcolor[rgb]{ .992,  .922,  .518}132\% & \cellcolor[rgb]{ .992,  .922,  .518}159\% \\
    \midrule
    \textbf{RP} & \cellcolor[rgb]{ .996,  .863,  .506}100\% & \cellcolor[rgb]{ 1,  .922,  .518}14\% & \cellcolor[rgb]{ .996,  .871,  .506}82\% & \cellcolor[rgb]{ .996,  .863,  .506}92\% & \cellcolor[rgb]{ .996,  .855,  .502}97\% & \cellcolor[rgb]{ .996,  .867,  .506}108\% & \cellcolor[rgb]{ .996,  .871,  .506}128\% \\
    \midrule
    \textbf{MDP} & \cellcolor[rgb]{ .996,  .922,  .518}117\% & \cellcolor[rgb]{ .984,  .914,  .514}14\% & \cellcolor[rgb]{ .996,  .922,  .518}94\% & \cellcolor[rgb]{ .996,  .922,  .518}108\% & \cellcolor[rgb]{ .992,  .922,  .518}117\% & \cellcolor[rgb]{ .996,  .922,  .518}124\% & \cellcolor[rgb]{ 1,  .922,  .518}144\% \\
    \midrule
    \textbf{Sust\_1} & \cellcolor[rgb]{ .973,  .42,  .42}35\% & \cellcolor[rgb]{ .569,  .796,  .49}10\% & \cellcolor[rgb]{ .973,  .412,  .42}22\% & \cellcolor[rgb]{ .973,  .412,  .42}27\% & \cellcolor[rgb]{ .973,  .424,  .42}35\% & \cellcolor[rgb]{ .973,  .412,  .42}40\% & \cellcolor[rgb]{ .973,  .42,  .42}54\% \\
\bottomrule
    \end{tabular}}%
  \label{tab:NASDAQ100_ROI_q3}%
\end{table}%
%%%%%%%%%%%%%%%%%%%%%%%%%%%%%%%%%%%%%%%%%%%%%%%%%%%%%%%%%%%%%%%%%%%%%%%%%%%%%%%%%%%%%%%%%%%%%%%%%%%%%%%%%%%%%%%%%%%%%%%%%%%%
% Table generated by Excel2LaTeX from sheet 'ROI750 (2)'
\begin{table}[htbp]
  \centering
  \caption{ROI based on a 3-year time horizon (with $k = 4$)}
    \scalebox{0.7}{\begin{tabular}{|c|c|c|c|c|c|c|c|}
    \toprule
    \textbf{Approach} & \textbf{ExpRet} & \textbf{Vol} & \textbf{5\%-perc} & \textbf{25\%-perc} & \textbf{50\%-perc} & \textbf{75\%-perc} & \textbf{95\%-perc} \\
    \midrule
    \textbf{MV0} & \cellcolor[rgb]{ .973,  .412,  .42}34\% & \cellcolor[rgb]{ .388,  .745,  .482}9\% & \cellcolor[rgb]{ .973,  .412,  .42}22\% & \cellcolor[rgb]{ .973,  .412,  .42}26\% & \cellcolor[rgb]{ .973,  .412,  .42}33\% & \cellcolor[rgb]{ .973,  .412,  .42}40\% & \cellcolor[rgb]{ .973,  .412,  .42}52\% \\
    \midrule
    \textbf{EW} & \cellcolor[rgb]{ .992,  .922,  .518}121\% & \cellcolor[rgb]{ 1,  .918,  .518}18\% & \cellcolor[rgb]{ .988,  .918,  .518}100\% & \cellcolor[rgb]{ .992,  .922,  .518}109\% & \cellcolor[rgb]{ .992,  .922,  .518}116\% & \cellcolor[rgb]{ .988,  .922,  .518}132\% & \cellcolor[rgb]{ .992,  .922,  .518}159\% \\
    \midrule
    \textbf{RP} & \cellcolor[rgb]{ .996,  .863,  .506}100\% & \cellcolor[rgb]{ .839,  .875,  .506}14\% & \cellcolor[rgb]{ .996,  .871,  .506}82\% & \cellcolor[rgb]{ .996,  .863,  .506}92\% & \cellcolor[rgb]{ .996,  .855,  .502}97\% & \cellcolor[rgb]{ .996,  .867,  .506}108\% & \cellcolor[rgb]{ .996,  .871,  .506}128\% \\
    \midrule
    \textbf{MDP} & \cellcolor[rgb]{ .996,  .922,  .518}117\% & \cellcolor[rgb]{ .812,  .867,  .506}14\% & \cellcolor[rgb]{ .996,  .922,  .518}94\% & \cellcolor[rgb]{ .992,  .922,  .518}108\% & \cellcolor[rgb]{ .992,  .922,  .518}117\% & \cellcolor[rgb]{ .996,  .922,  .518}124\% & \cellcolor[rgb]{ 1,  .922,  .518}144\% \\
    \midrule
    \textbf{Sust\_1} & \cellcolor[rgb]{ .973,  .42,  .42}35\% & \cellcolor[rgb]{ .427,  .757,  .482}10\% & \cellcolor[rgb]{ .973,  .412,  .42}22\% & \cellcolor[rgb]{ .973,  .412,  .42}27\% & \cellcolor[rgb]{ .973,  .424,  .42}35\% & \cellcolor[rgb]{ .973,  .412,  .42}40\% & \cellcolor[rgb]{ .973,  .42,  .42}54\% \\
\bottomrule
    \end{tabular}}%
  \label{tab:NASDAQ100_ROI_q4}%
\end{table}%
%
%%%%%%%%%%%%%%%%%%%%%%%%%%%%%%%%%%%%%%%%%%%%%%%%%%%%%%%%%%%%%%%%%%%%%%%%%%%%%%%%%%%%%%%%%%%%%%%%%%%%%%%%%%%%%%%%%%%%%%%%%%%%%%%%
\end{document}